\documentstyle[12pt]{article}

\begin{document}
\def\tfrac#1#2{{\textstyle{#1\over #2}}}
\def\bch{Q}
\def\Qop{{\bf Q}}
\def\Vop{{\bf V}}
\def\p{\partial\phi_2}
\def\pp{\partial^2\phi_2}
\def\ppp{\partial^3\phi_2}
\def\bph{{\bf \varphi}}
\def\brh{{\bf \rho}}
\def\be{\begin{equation}}
\def\ee{\end{equation}}
\def\ba{\begin{eqnarray}}
\def\ea{\end{eqnarray}}
\def\dif{\partial}

\pagestyle{empty}
\rightline{UG-7/93}
\rightline{ITP-SB-93-79}
\rightline{}
\vspace{1.1truecm}
\centerline{\bf  On the Cohomology of the Noncritical $W$-string}
\vspace{1.1truecm}
\centerline{E.~Bergshoeff ${}^*$,
J.~de Boer ${}^\dagger$, M.~de Roo ${}^*$ and T.~Tjin ${}^\ddagger$ }
\vspace{1.1truecm}
\noindent $^*$ {\sl Institute for Theoretical Physics,
Nijenborgh 4, 9747 AG Groningen, The Netherlands.}

\vspace{.5truecm}

\noindent $^\dagger$ {\sl Institute for Theoretical Physics, State University
of New York, Stony Brook, NY 11794-3840, USA.}

\vspace{.5truecm}

\noindent $^\ddagger$
{\sl University of M\"unchen, Theresienstr.\ 37, D-80333 M\"unchen,
 Germany}

\vspace{1.5truecm}

\centerline{ABSTRACT}

\vspace{.5truecm}

We investigate the cohomology structure of a general noncritical
$W_N$-string. We do this by introducing a new basis in the
Hilbert space in which the BRST operator splits into a ``nested''
sum of nilpotent BRST operators. We give explicit details for the
case $N=3$. In that case the BRST operator $Q$ can be written as the sum
of two, mutually anticommuting, nilpotent BRST operators: $Q=Q_0+Q_1$.
We argue
 that if one chooses for the Liouville sector a $(p,q)$ $W_3$
minimal model then the cohomology of the $Q_1$ operator is closely related
to a $(p,q)$ Virasoro minimal model. In particular, the special case of a
(4,3) unitary $W_3$ minimal model with central charge $c=0$
leads to a $c=1/2$ Ising model in the $Q_1$ cohomology. Despite all this,
noncritical $W_3$ strings are not identical to noncritical Virasoro strings.

\vfill\eject
\pagestyle{plain}
\noindent{\bf 1. Introduction}

\vspace{.5cm}
Noncritical strings are strings in which the two-dimensional
gravitational fields do not decouple after quantization but instead
develop an induced kinetic term. The string coordinates are
called ``matter'' fields while the
non-de\-coup\-led gravity fields are represented by a set of so-called
``Liouville'' fields. The matter and Liouville fields together
form a realization of the Virasoro algebra. The spectrum of the
noncritical string in less than or precisely one dimension
has been calculated via a BRST analysis in \cite{li1,li1a,li1b}.

Unfortunately, if the number of dimensions in which the noncritical
string propagates is larger than one, tachyons appear in the spectrum,
and the theory is ill-defined. It is believed that it is possible to
make sense out of noncritical strings in more than one dimension,
provided the underlying Virasoro algebra is extended to a
nonlinear so-called $W$-algebra \cite{de1}.
In the last few years much research has been devoted to investigating
the structure of these $W$-symmetries
(for a recent review, see e.g.~\cite{sc1}).
Noncritical string
theories with an underlying $W$-symmetry are referred to as
noncritical $W$-strings.

One expects that in a noncritical $W$-string the matter and Liouville
fields (representing the ``$W$-gravity'' sector) separately form a
realization of the underlying nonlinear $W$-algebra.
Thus we are faced with an
immediate problem: due to the nonlinear nature of the $W$-algebra it
is clear that the matter and Liouville fields {\sl together} do not
realize the same $W$-algebra. This would seem to make the construction of a
nilpotent BRST operator problematic. A way out of this apparent obstacle
was provided by the work of \cite{ber1} where it was shown that nevertheless
a nilpotent BRST operator for the matter + Liouville system could be
constructed. This is made possible by the fact that at the classical level
the sums of the generators of the $W$ algebras in the
matter and Liouville sectors still form a closed Poisson bracket algebra,
albeit with field-dependent structure functions \cite{be1}. For
definiteness, we will call this algebra the ``modified''
${\tilde w}$-algebra\footnote{We denote the quantum extension of
a classical $w$-algebra, provided it exists, by a $W$-algebra.}. Alternatively,
the existence of this BRST operator can be understood from the covariant
action for $W$ gravity coupled to matter \cite{bg}.

Having a nilpotent BRST operator at one's disposal one can proceed with
a calculation of its cohomology and thus the spectrum of the noncritical
$W$-string. Some results in this  direction have been obtained in
\cite{ber1,ber2,bo1,bo4,bo1a}.

Recently, it has been shown that the BRST operator of an ``unmodified''
$W_3$ algebra\footnote{
This case is often referred to as the ``critical''
 $W$-string. The possibility of constructing new critical string
 theories by exploiting the $W$-symmetries was first suggested
 in \cite{bil1}. Note that
 the ``critical'' $W$-string can be obtained from the
``noncritical'' $W$-string by setting the Liouville fields equal to zero.}
can be decomposed as the sum of two separate, mutually
anticommuting, BRST operators by performing a particular canonical
transformation \cite{po1}. This is related to the fact that
 after the canonical transformation
 the spin-three generators of the {\sl classical}
 $w_3$ algebra form a closed Poisson bracket algebra with
 field-dependent structure constants \cite{be2}. It turns
 out that quite generally the BRST operator corresponding
 to a $W_N$ algebra can be written as a ``nested'' sum of
 nilpotent BRST operators
 and the same applies to the BRST operator of the noncritical $W$-string
 \cite{be3}.

One of the advantages of decomposing a BRST operator into a ``nested''
sum of nilpotent BRST operators, is that one can now construct the
cohomology by an iterative procedure. One first calculates the
cohomology of one of the BRST operators occurring in the sum.
In practice it is easiest to start with the BRST operator corresponding
to the highest-spin generator (see \cite{be3} for more details).
One then includes the next to highest-spin generator and the cohomology
of the combined system can be obtained from the one corresponding to
the highest
spin generator by a so-called ``tic-tac-toe'' procedure \cite{bot1}.

In \cite{da1,ra1}, a relationship has been suggested, on the basis of
comparing values of central charges, between the spectra of
``critical'' $W_N$-strings and Virasoro minimal models. In the
case of the $W_3$-string this relation has been made more explicit in
\cite{po1,po2,we1,we2,hu1}. In particular, it was shown that a general physical
state of the critical $W_3$-string contains a factor corresponding to
a $c=1/2$ Ising model. In terms of the new basis discussed above,
 it means that the cohomology  of the
 BRST-operator of the highest-spin symmetry corresponds
 to a $c=1/2$ Ising model.

Using  similar arguments as in \cite{da1,ra1}, it can be shown
that the corresponding situation for the noncritical $W_N$-string is
as follows \cite{be3}. Given a realization of the matter and Liouville
sectors in terms of $N-1$ free scalars, one can always perform a
canonical transformation in the matter sector\footnote{The discussion
below can be repeated if one performs the canonical transformation in
the Liouville sector.} such that the
``modified'' ${\tilde w}_N$-algebra manifestly
has a ``nested'' set of subalgebras
\begin{equation}
v_N^N \subset v_N^{N-1}\subset \cdots \subset v_N^2 \equiv {\tilde w}_N\,,
\end{equation}
where the subalgebra $v_N^n$ consists of generators of spin $s=\{n,
n+1,\cdots,N\}$, respectively. Each generator of spin $s=n$ depends
on $N-n+1$ of the $N-1$ matter scalars and all the $N-1$
Liouville scalars. In the new basis the BRST charge $Q_N$ of the
${\tilde w}_N$-algebra has the following nested structure:
\begin{equation}
\label{eq:Qincl}
Q_N^N \subset Q_N^{N-1} \subset \cdots Q_N^2 \equiv Q_N\,,
\end{equation}
where $Q_N^n$ is the BRST charge corresponding to the subalgebra
$v_N^n$. The BRST charge $Q_N^n$ depends on $N-n+1$ matter scalars,
all the $N-1$ Liouville fields and the ghost and anti-ghost fields of the
spin $n,\cdots,N$ symmetries.
The inclusion symbols in (\ref{eq:Qincl}) indicate that the BRST charge
$Q_N^n$ can be obtained from the BRST charge $Q_N^{n-1}$ by
setting in the expression for  $Q_N^{n-1}$ the ghosts and antighosts
corresponding to the spin-$(n-1)$ symmetries equal to zero.
In all explicit examples considered thus far
the nested structure of the BRST charges (\ref{eq:Qincl})
survives quantization, where
the BRST charges become BRST operators.
To distinguish between BRST operators and
BRST charges, we will write the operators with boldface\footnote{
The explicit form of the
${\bf Q}_N^N$ operator for $N=4,5,6$ (with the Liouville fields set equal
to zero)
is given in \cite{po3} while the explicit expression for the ${\bf Q}_4^3$
operator
(without zero Liouville fields) can be found in \cite{be3}.
 In the usual Miura basis
 \cite{fa1,da1}, the BRST operator of the $W_4$-algebra was given in
 \cite{horn1,zhu1}.}.
We note that the case $N=3$
is special. In that case the BRST operator decomposes as the
sum of two, mutually anticommuting, nilpotent BRST operators
as follows:

\begin{equation}
\label{eq:W3}
{\bf Q}_3 = {\bf Q}_0 +{\bf Q}_1\,,
\end{equation}
where ${\bf Q}_1 = {\bf Q}_3^3$ and ${\bf Q}_0$ is the nilpotent
BRST operator corresponding to the Virasoro subalgebra.

By comparing the values of central charges, it can now be argued
that, if one restricts the Liouville sector of a noncritical
$W_N$-string to a $(p,q)$ $W_N$ minimal model, then the cohomology
of the $Q_N^n$ BRST operator is closely related to the
$(p,q)$ minimal model of the $W_{n-1}$-algebra.

More precisely, the counting of the central charges is as follows \cite{be3}.
Both for the matter and the Liouville sectors we realize the $W_N$ algebra
by $N-1$ scalars $\phi_k$ and $\sigma_k$, respectively. The background
charges of these fields are fixed by the Miura transformation \cite{fa1,da1}.
Consider now the ${\bf Q}_N^n$
operator. It
depends on $N-n+1$ matter scalars, the Liouville fields and the ghosts and
antighosts for the spin $n,\cdots,N$ symmetries. In general, the central charge
contribution of the spin $n,\cdots,N$ ghosts and antighosts for a given
value of $n$ is given by
\begin{equation}
c^n_{gh} =  -2 \sum_{k=n}^N (6k^2-6k+1)\,.
\end{equation}
In particular, the contribution of {\it all} ghosts and antighosts is given by
\begin{equation}
c_{gh} \equiv c_{gh}^2 = -2 (N-1)(2N^2+2N+1)\,.
\end{equation}
The central charge contributions of the Liouville and
matter sector are given by
\begin{eqnarray}
c_l &=& (N-1)\bigg \{1 - 4\big (Q^2-N(N-1)\big ){N+1\over N-1}\bigg \}\,,\cr
c_m &=& (N-1)\bigg \{1 + 4Q^2 {N+1\over N-1}\bigg \}\,,
\end{eqnarray}
where $Q$ is a free parameter which can be identified with the
background charge of the matter field $\phi_{N-1}$.
Note that we have
\begin{equation}
c_m + c_l + c_{gh} = 0\,,
\end{equation}
as is required to allow for a nilpotent BRST operator.
We now choose for $Q$ one of the following values
\begin{equation}
\label{eq:pqmin}
Q_{min}^2 = N(N-1){(p+q)^2\over 4pq}\,,
\end{equation}
where $p$ and $q$ are non-negative integers which are relatively prime.
The central charge contribution of the Liouville sector is then given by
\begin{equation}
c^{(p,q)}_l = (N-1)\bigg\{1-{(p-q)^2\over pq}N(N+1)\bigg\}\,,
\end{equation}
which corresponds to the $(p,q)$ minimal model of the $W_N$-algebra.
Similarly, the central charge contribution of the matter sector is given by
\begin{equation} \label{jj1}
c^{(p,q)}_m = (N-1)\bigg\{1+{(p+q)^2\over pq}N(N+1)\bigg\}\,,
\end{equation}
which corresponds to the $(p,-q)$ minimal model of the
$W_N$ algebra. Finally, for fixed $n$,
the central charge contribution of the $N-n+1$ matter
scalars is given by
\begin{equation}
c^n_m = \sum_{k=n-1}^{N-1} \big (1+12(\alpha_k)^2\big )\,,
\end{equation}
where $\alpha_k$ is the background charge of the matter scalar
$\phi_k$:
\begin{equation}
\alpha_k = Q_{min}{\sqrt {k(k+1)\over N(N-1)}}\,.
\end{equation}
Denoting all the different contributions of the fields occurring
in the ${\bf Q}_N^n$ operator by $c_N^{n;(p,q)}$
we find for a given choice of
$p,q$ and $n$
\begin{eqnarray}
c_N^{n;(p,q)} &=& c^{(p,q)}_l + c^n_m + c^n_{gh} \cr
&=& (n-2)\bigg\{(2n-1)^2 - n(n-1) {(p+q)^2\over pq}\bigg\}\,,
\end{eqnarray}
which corresponds to the $(p,q)$ minimal model of the $W_{n-1}$ algebra.
This concludes our counting argument.

An interesting special case arises if we choose the Liouville
sector to correspond to the $(N+1,N)$ $W_N$ minimal model with
$c_l^{(N+1,N)}=0$
\cite{be2}. In that case the Liouville fields effectively decouple
from the theory
and we end up with a ``critical'' $W_N$ string. The central charge
corresponding to the ${\bf Q}_N^n$ operator is in that case given by
\begin{equation}
c_N^{n;(N+1,N)} = (n-2)\bigg \{1-{n(n-1)\over N(N+1)} \bigg\}\,,
\end{equation}
which is the $(N+1,N)$ minimal model of the $W_{n-1}$ algebra\footnote{
This relation between critical $W$-strings and minimal models has been
suggested in \cite{po4,hu1}.}.
For instance, for $n=N=3$ we find the $c_3^{3;(4,3)} = 1/2$ Ising model.

Another interesting case occurs if we choose the Liouville sector
to correspond to the $(n,n-1)$ minimal model of the $W_N$ algebra
with central charge  ($2 \le n \le N)$
\begin{equation}
\label{eq:cl}
c_l^{(n,n-1)} = (N-1) \bigg\{ 1 - {N(N+1)\over n(n-1)}\bigg\}\,.
\end{equation}
In that case we find that $c_N^{n;(n,n-1)} = 0$. This means that all
fields occurring in the ${\bf Q}_N^n$ cohomology decouple and we
effectively end up with a ``critical'' $W_{n-1}$-string theory.
An interesting additional feature of the Liouville central charge
given in (\ref{eq:cl}) is that for this value
the total central charge contribution of the $n-2$ Liouville scalars
$\sigma_1,\cdots,\sigma_{n-2}$ equals zero. For instance, if $n=N$
then $c_l^{(N,N-1)}=-2$ and the central charge contribution of
the $N-2$ Liouville scalars $\sigma_1,\cdots,\sigma_{N-2}$ equals zero.
In the particular case $N=3$ \cite{be1},
we have two Liouville scalars $\sigma_1$ and
$\sigma_2$. The central charge contribution of $\sigma_1$ is zero
and the whole $W_3$ algebra can in fact be realized in terms of the
{\sl single} scalar $\sigma_2$ \cite{bo2}.

 From a group-theoretic point of view, the picture is the following. The
matter fields can be seen as one scalar field $\bph$ with values in the Cartan
subalgebra of $sl_{N}$. The energy momentum tensor of these fields
 can be written as
\begin{equation} \label{jj2}
T=-\frac{1}{2} \partial \bph \cdot \partial \bph + B\,  \brh_N \cdot
\partial^2 \bph\,,
\ee
 with central charge $c=(N-1)+12\, B^2 \brh_N\cdot\brh_N$. Here,
 $\brh_N$ is half the sum
 of the positive roots of $sl_N$.
 If the Liouville sector is a $(p,q)$ minimal
 model, then one finds, on using (\ref{jj1}) and
 $\brh_N\cdot\brh_N=(N^3-N)/12$, that $B=\sqrt{p/q}+\sqrt{q/p}$\footnote{
 The parameter $B$ is related to $Q_{min}$ (\ref{eq:pqmin})
 by $Q_{min}=\tfrac{1}{2}\sqrt{N(N-1)}\,B$.}.
 To get the $\Qop^n_N$ operator, we have to
 decouple $n-2$ matter fields, which can be done as follows. Take an
 embedding of $sl_{n-1}$ into $sl_N$, for instance map the $n-2$ simple
 roots of $sl_{n-1}$ to the first $n-2$ simple roots of $sl_N$. Then, decompose
 the field $\bph$ as $\bph_{n}+\bph^{\perp}$, where $\bph_{n}$ is the
 orthogonal projection of $\bph$ onto the Cartan subalgebra of $sl_{n-1}$.
 The $n-2$ matter fields we want to decouple are precisely given by
 the components of $\bph_n$. To find the energy momentum tensor for $\bph_n$,
 we have to decompose $\brh_N$ into a piece with values in the $sl_{n-1}$
 Cartan subalgebra and a piece perpendicular to that. But this latter piece is
 precisely
 $\brh_N-\brh_{n-1}$, because both $\brh_N$ and $\brh_{n-1}$ have inner
 product one with the simple roots $\alpha_1,\ldots,\alpha_{n-2}$, so that
 their difference has inner product zero with respect to these simple roots
 and is therefore perpendicular to the $sl_{n-1}$ Cartan subalgebra.
 This demonstrates that the energy momentum tensor for $\bph_n$ is identical
 to (\ref{jj2}), but with $\brh_N$ replaced by $\brh_{n-1}$. The value
 of $B$ remains the same, so that the central charge of the fields $\bph_n$
 is equal to central charge of the matter fields in a theory of noncritical
 $W_{n-1}$ gravity, where the Liouville fields form a $(p,q)$ minimal model
 for the $W_{n-1}$ algebra. This is the same statement as we made previously,
 and we see that the fields that enter the $\Qop^n_N$ operator are those
 orthogonal to an $sl_{n-1}$ subalgebra of $sl_N$. Later, we will use this
 observation to try to improve our understanding of  the results of the
 computation of the $\Qop_3^3\equiv \Qop_1$ cohomology.

Since the BRST operator for the noncritical $W_N$ string can be derived
via Hamiltonian reduction from the superalgebra $sl(N|N-1)$, it is an
interesting idea to try derive the operators ${\bf Q}^n_N$ from Hamiltonian
reduction as well, employing the group-theoretical insight obtained above.

In the above discussion, all relations between noncritical
$W$ strings and minimal models
have been based upon comparing values of central charges. It
is the purpose of this paper to confirm the relations suggested by
the above counting arguments for $N=3$
by considering the weights of the primary fields
occurring in the cohomology.
We will verify that these weights indeed correspond to the
relevant minimal model. Note that for $N=3$ the BRST operator
decomposes according to (\ref{eq:W3}).
Our main result can be stated as follows.
Starting from a $(p,q)$ $W_3$ minimal model in the Liouville sector,
we find that the primary fields occurring in the cohomology
of the ${\bf Q}_1$ operator exactly correspond to
those of the $(p,q)$
minimal model of the Virasoro algebra. More precisely, all but one
of the primary fields occur as tachyonic states at level 0 while
the remaining one occurs at level one.
We suggest that all other solutions of the cohomology can be obtained
by the action of so-called picture-changing and screening operators
on this basis set of primary fields.

The organization of the paper is as follows. In section 2
we introduce some basic notions of the noncritical $W_3$-string.
In section 3
we consider the primary fields occurring in the ${\bf Q}_1$
cohomology and show that for the choice (\ref{eq:pqmin})
of the background charge $Q$
indeed they correspond to $(p,q)$ minimal
models of the Virasoro algebra. In section 4 we
consider the screening operators and the picture changing operators
of the ${\bf Q}_1$ operator. We will argue that the complete cohomology of
${\bf Q}_1$ is obtained by acting on the basic primary fields with
strings of screening and picture changing operators\footnote{
This concerns the states with continuous momenta. The situation
of the discrete states is more complicated.}.
Next, in section 5 we illustrate our results by
working out examples for specific values of $p$ and $q$.
Finally, in section 6 we will discuss the cohomology of the
${\bf Q}_1$ operator for a $W_3$ minimal model, the $\Qop_0+\Qop_1$
cohomology and generalizations to other $W$ algebras.
In particular we discuss in what way noncritical Virasoro strings are
embedded in noncritical $W_3$ strings. Our results indicate that
the the $W_3$ $(p,q)$ minimal model coupled to $W_3$ gravity is not
the same theory as the Virasoro $(p,q)$ minimal model coupled to
ordinary gravity, although it is contained in the noncritical
$W_3$ string. The Virasoro minimal model is recovered by taking
a modified noncritical $W_3$ string, that contains an additional
set of screening operators. It is interesting to compare this
with the work of Berkovits and Vafa \cite{beva}, who demonstrated how
special $N=i$ strings reduce to arbitrary $N=j$ strings, $j<i\leq2$.
It is tempting to conjecture that strings based on certain chiral
algebras can always somehow be embedded in strings based on larger
chiral algebras. For example, it was shown recently
\cite{bfw} that the ordinary
bosonic string can be realized as a specific critical $W_3$ string.
This realization differs from ours, since we realize certain bosonic
strings as subsectors of non-critical $W_3$ strings. It would be
interesting to see if there is a deeper relation between the two
realizations.

\vspace{.5truecm}

\noindent{\bf 2. The noncritical $W_3$-string}
\vspace{.5cm}

In this section we will illustrate the general structure of the
quantum BRST operators $\Qop_N^n$
for the noncritical $W_N$-string with the $N=3$ case. We will employ
a simple realisation of the noncritical $W_3$-string in terms of
two scalars in the matter sector, as well as two scalars for the
Liouville or $W$-gravity sector, two scalars being the minimum
number for a realisation of the $W_3$-algebra with arbitrary central
charge.

In Table 1 we present these fields with the corresponding background
 and central charges. These charges are limited by the requirement that the
 total central charge equals 100, thus cancelling the contribution
 from the spin-2 and spin-3 ghosts,
 and by the fact that the matter
 and Liouville sectors separately realise a $W_3$-algebra.
 This fixes the relative coefficient between the background
 charges in these two-scalar realisations \cite{ro1}.

In terms of roots of $sl_3$, $\phi_1$ corresponds to $\alpha_1$ and
$\phi_2$ to $2\alpha_2+\alpha_1$.

In \cite{be2} the BRST-operator of the noncritical $W_3$-string
in the new basis, discussed in the introduction, was calculated.
It is given by $\Qop=\oint j$, with $j=j_0+j_1$ given by

\vspace{.25cm}
\begin{center}
\renewcommand{\arraystretch}{1.5}
\begin{tabular}{|l||l|l|}
\hline
field  &\hfil background charge\hfil&\hfil central charge\hfil\\
\hline
$\phi_1$&$\tfrac{1}{3}\sqrt{3}\,{\bch}$&$1+4{\bch}^2$\\
$\phi_2$&${\bch}$&$1+12{\bch}^2$\\
\hline
$\sigma_1$&$\tfrac{1}{3}\sqrt{3(6-{\bch}^2)}$&$25-4{\bch}^2$\\
$\sigma_2$&$\sqrt{6-{\bch}^2}$&$73-12{\bch}^2$\\
\hline
$c,\ b$&\hfil&$-26$\\
$\gamma,\ \beta$&\hfil&$-74$\\
\hline
\end{tabular}
\renewcommand{\arraystretch}{1.0}
\end{center}
\vspace{.25truecm}

\noindent {\bf Table\ 1.} \ \ \ \ \ The fields of
the noncritical $W_3$-string. The scalar fields of the matter
sector ($\phi_1,\ \phi_2$) and of the Liouville sector
($\sigma_1, \sigma_2$) are given
with their background charge, and their contribution
to the central charge. The fields ($c,\ b$) are the spin-2,
($\gamma,\ \beta$) the spin-3 ghosts.
\vspace{.5truecm}

\begin{eqnarray}
\label{J0}
 j_0 &=& c\,\{ T_M + T_L
     + T_{(\gamma,\beta)} +{1\over 2} T_{(c,b)} \} \,,
  \\
 j_1 &=&\gamma\,\big[\,
  {i\over 3\sqrt{6}}\,\{
   4(\p)^3 - 12{\bch}\p\pp + (-15+4{\bch}^2)\ppp \}
  \nonumber\\
  &&\qquad + i\,\{W_L- {2\over \sqrt{6}}\p T_L +{1\over \sqrt{6}}{\bch}
  \partial T_L\}
   \nonumber\\
  \label{J1}
  &&\quad -i\sqrt 6\{ \p\partial\gamma\beta
+{1\over 3}{\bch}\partial\beta\partial\gamma\}\big] \,.
\end{eqnarray}
The main advantage of the chosen basis is that $\Qop_0=\oint j_0$ and
 $\Qop_1=\oint j_1$ are seperately nilpotent, and therefore anticommute.

Note that the scalar  $\phi_1$, and the spin-2 ghosts are
 absent from $j_1$, and that the Liouville fields do not occur
 explicitly in (\ref{J1}), but only in
\begin{eqnarray}
  T_L&=& T_{\sigma_1}+T_{\sigma_2} \nonumber\\
\label{TL}
     &=& -\tfrac{1}{2}(\partial\sigma_1)^2 +
            \tfrac{1}{3}\sqrt{3(6-{\bch}^2)}\,
            \partial^2\sigma_1
         -\tfrac{1}{2}(\partial\sigma_2)^2 + \sqrt{6-{\bch}^2}\,
            \partial^2\sigma_2\,,\\
  W_L&=& \tfrac{1}{18}i\sqrt{6}\bigg\{
     (\partial\sigma_2)^3
    - 3\sqrt{6-{\bch}^2}\,\partial\sigma_2\partial^2\sigma_2 \nonumber\\
\label{WL}
   &&\qquad
    +(6-{\bch}^2)\partial^3\sigma_2 +
    6 \partial\sigma_2 T_{\sigma_1}
    -3\sqrt{6-{\bch}^2}\,\partial T_{\sigma_1}
     \bigg\} \,.
\end{eqnarray}
$T_L$ and $W_L$ satisfy a $W_3$ algebra. In $j_0$ we find, besides $T_L$,
 the energy-momentum tensors of the matter and ghost sectors:
\begin{eqnarray}
  T_M&=& T_{\phi_1}+T_{\phi_2} \nonumber\\
\label{TM}
     &=& -\tfrac{1}{2}(\partial\phi_1)^2 +
            \tfrac{1}{3}\sqrt{3}{\bch}
            \partial^2\phi_1
         -\tfrac{1}{2}(\partial\phi_2)^2 + {\bch}
                  \partial^2\phi_2\,,\\
\label{Tgh2}
  T_{(c,b)} &=& -2 b \partial c - (\partial b) c \,,\\
\label{Tgh3}
  T_{(\gamma,\beta)}&=&  -3 \beta \partial \gamma -
      2 (\partial \beta) \gamma \,.
\end{eqnarray}

For the convenience of the reader we will give some of the formulae
 of the introduction specifically for $N=3$.
 The total Liouville central charge $c_l$, and $c_3^3$, the
 contribution to the central charge of
 the fields that play a role in $\Qop_1$ ($\phi_2$, the spin-3 ghosts,
 and the Liouville scalars) are give by:
\begin{eqnarray}
\label{cL}
  c_l&=&98-16{\bch}^2 \,, \\
\label{c1}
  c_3^3&=&25-4{\bch}^2\,.
\end{eqnarray}
If we choose ${\bch}$ equal to
\begin{eqnarray}
\label{Qmin}
 {\bch}_{min}&=&{3(q+p)\over \sqrt{6pq}} \,,\ \
  {\sqrt { 6 - {\bch}_{min}^2}} =
  {3i(q-p)\over \sqrt{6pq}}
  \, .
\end{eqnarray}
 we find the following values for $c_l$ and $c_3^3$:
\begin{eqnarray}
\label{cminL}
   c_l^{(p,q)} &=& 2(1-{12(p-q)^2\over pq})\,, \\
\label{cmin1}
  c_3^{3;(p,q)} &=& 1 - {6(p-q)^2\over pq} \,.
\end{eqnarray}
The values $c_l^{(p,q)}$ and $c_3^{3;(p,q)}$ correspond to the central charges
 of the $(p,q)$ $W_3$ minimal model, and the $(p,q)$ Virasoro minimal
 model, respectively. This relationship between $W_3$ and Virasoro
 minimal models will be further elucidated in the next two sections, where
 we will consider the cohomology of $\Qop_1$.

With the above formulae for $N=3$, it is a simple matter to verify the
 relations between central charges in the special cases considered
 in the introduction. After treating the $\Qop_1$-cohomology in the
 next two sections, we will come back to these examples in Section 5.

\vspace{.5cm}
\noindent{\bf 3. The $\Qop_1$ cohomology}
\vspace{.5cm}

In this section we will consider the cohomology of the operator
 $\Qop_1$, acting on Fock spaces of $\phi_2,\sigma_1,\sigma_2$ and
of the ghosts $\beta,\gamma$. The momenta of $\sigma_1$ and $\sigma_2$
will be chosen to be equal to those appearing in the Felder resolution
of a $W_3$ minimal model. In section~6 we explain the implications of
these results for the computation of the $\Qop_1$ cohomology when the
Fock space of $\sigma_1,\sigma_2$ is replaced by the Hilbert space of a
$W_3$ minimal model.
In particular, for the values of the background charge
${\bch}$ corresponding
 to a $W_3$ minimal model in the Liouville sector, and the momenta
of $\sigma_1,\sigma_2$ equal to those in the corresponding Felder
resolution, we want to
 identify the primary operators of the Virasoro  minimal
 model with central charge (\ref{cmin1}).

For this it turns out to be
 sufficient to consider
 the states
 at level 0 and level 1.
 At level 0 the states of lowest ghost number are of the
 form\footnote{The level of a state is defined by ${\rm level} \equiv h+3$,
 where $h$ is the weight of the fields in front of the exponential.
 For instance, the level of the state (\ref{vac}) is given by
 (-1-2)+3=0. The state (\ref{vac}) is the state of lowest ghost
 number at level zero. Other level zero states, with ghost number
 one higher, have a factor $(\partial^2\gamma)(\partial\gamma)\gamma$
 in front of the exponential.}:
\begin{eqnarray}
\label{vac}
  \Vop_0(p_2,s_1,s_2) &=& (\partial{\gamma})\gamma\, {\rm e}^{ip_2\phi_2+
             is_1\sigma_1+is_2\sigma_2} \,.
\end{eqnarray}
The condition $\Qop_1 \Vop_0(p_2,s_1,s_2)=0$ determines the momenta of
 the three fields. The resulting
 cubic equation factorizes, and we obtain the
 following three solutions:
\begin{eqnarray}
\label{vacA}
  (A_0)&&p_2=is_2 -i{\bch} -\sqrt{6-{\bch}^2}\,,\\
\label{vacB}
  (B_0)&&p_2=+\tfrac{1}{2}i\sqrt{3}s_1 -\tfrac{1}{2}is_2-i{\bch} \,,\\
\label{vacC}
  (C_0)&&p_2=-\tfrac{1}{2}i\sqrt{3}s_1 -\tfrac{1}{2}is_2-i{\bch}
    +\sqrt{6-{\bch}^2} \,,
\end{eqnarray}
where $s_1$ and $s_2$ are arbitrary.

Now, if we choose the parameter $Q$ equal to (\ref{Qmin}), the
 central charge of the Liouville sector corresponds to that of a
 $(p,q)$ $W_3$ minimal model. By choosing $s_1$ and $s_2$ appropriately,
 we restrict these momenta to those of this minimal model, and in that
 case we expect that the states determined by (\ref{vacA}-\ref{vacC})
 should correspond to the primary states of the $(p,q)$ minimal
 Virasoro model. The allowed values of $s_1$ and $s_2$ are  (see, e.g.,
 \cite{sc1})
\begin{eqnarray}
\label{s1val}
  s_1 &=& -{1\over \sqrt{2pq}}(qr_2-pt_2) \,,\\
\label{s2val}
  s_2 &=& -{1\over \sqrt{6pq}}
     (2(qr_1-pt_1) + (qr_2-pt_2)) \,,
\end{eqnarray}
with the non-negative integers $r_1,r_2,t_1,t_2$ restricted
 according to $0\le r_1+r_2\le p-3,\ 0\le t_1+t_2\le q-3$.

For the above values of $p_2,s_1$ and $s_2$ we now calculate
 the conformal weight of the state (\ref{vac}).
 The conformal weight of a
 vertex operator $\exp{(ip\phi)}$, for a generic scalar $\phi$ with
 background charge $Q_{\phi}$, is given by
\begin{eqnarray}
     h_{\phi}&=& \tfrac{1}{2}\big( p+i Q_{\phi} \big)^2
                    + \tfrac{1}{2} Q_{\phi}^2 \,.
\end{eqnarray}
This, together with
 the weight $-3$ of the ghost contribution to (\ref{vac}),
 determines the allowed values of the conformal weights at level 0.

Note that the weight due to the Liouville fields is invariant  under
 the following transformation of the labels $r_1,r_2,t_1,t_2$
 \cite{sc1}:
\begin{eqnarray}
\label{rt-trans1}
  && (r_1,r_2,t_1,t_2) \to (p-3-r_1-r_2,r_1,q-3-t_1-t_2,t_1) \,.
\end{eqnarray}
Applying this transformation gives successively new Liouville momenta
 $(s_1^{\prime},s_2^{\prime})$ and $(s_1^{\prime\prime},s_2^{\prime\prime})$,
 while applying (\ref{rt-trans1}) three times gives back the original
 momenta. A second transformation which leaves the Liouville weights
 invariant but changes the sign of teh $W_3$-weight is
\begin{eqnarray}
\label{rt-trans2}
  && (r_1,r_2,t_1,t_2) \to (r_2,r_1,t_2,t_1) \,.
\end{eqnarray}
Note that $(r_1,r_2,t_1,t_2)$ cannot be invariant under (\ref{rt-trans1}).
 Therefore the weights of states with momenta (\ref{s1val}, \ref{s2val})
 are either six-fold
 or three-fold degenerate, the possibility of three-fold degeneracy
 occurring when the $W_3$-weight, which is proportional to
\cite{sc1}
\begin{eqnarray}
  &&(q(r_1-r_2)-p(t_1-t_2))(q(r_1+2r_2+3)-p(t_1+2t_2+3))\times
   \nonumber\\
  &&\qquad\qquad\times
   (q(2r_1+r_2+3)-p(2t_1+t_2+3))\,,
\end{eqnarray}
vanishes.
In a $W_3$ minimal model, representations whose
labels are related through (\ref{rt-trans1})
or (\ref{rt-trans2}) should be identified,
since the corresponding irreducible $W_3$ representations
are isomorphic. In a $W_3$ minimal model each of these should
occur with multiplicity one only.

The three solutions (\ref{vacA}-\ref{vacC}) are in fact related
 by the transformation (\ref{rt-trans1}). We find:
\begin{eqnarray}
  && p_2(s_1^\prime,s_2^\prime)(A_0) = p_2(s_1,s_2)(C_0) \,,\quad
  \nonumber\\
  && p_2(s_1^\prime,s_2^\prime)(B_0) = p_2(s_1,s_2)(A_0) \,,\quad
  \nonumber\\
  &&p_2(s_1^\prime,s_2^\prime)(C_0) = p_2(s_1,s_2)(B_0) \,.
\end{eqnarray}
Therefore the weights obtained for the three solutions
 (\ref{vacA}-\ref{vacC}) are the same. In the next section we will show
 that these relations between (\ref{vacA}-\ref{vacC}) can be
 understood from the action of certain screening operators.

The conformal weights of the primary operators of the $(p,q)$ Virasoro
 minimal model are given by
\begin{eqnarray}
\label{hvir}
   h_{Vir}(r,t)&=& {(q(r+1)-p(t+1))^2 - (q-p)^2 \over 4pq}\,,
\end{eqnarray}
for $0\le r\le p-2,\ 0\le t\le q-2$. $h_{Vir}$ is invariant under
 the change of labels
\begin{eqnarray}
\label{Virtrans}
   r\to p-2-r,\ t\to q-2-t\,.
\end{eqnarray}
Again, in a Virasoro minimal model, we should identify two representations
whose labels are related through (\ref{Virtrans}), since they give rise
to isomorphic irreducible Virasoro modules.

We should now identify the calculated weights of (\ref{vac}) with
(\ref{hvir}).
The following
Virasoro weights are obtained:
\begin{eqnarray}
\label{hvir0}
   (A_0) &\rightarrow& h_{Vir}(r_2,t_2)\,, \cr
   (B_0) &\rightarrow&   h_{Vir}(1+r_1+r_2,1+t_1+t_2)\,,\cr
   (C_0) &\rightarrow&  h_{Vir}(r_1,t_1)\,,
\end{eqnarray}
with the values of $r_1,r_2,t_1,t_2$ restricted by the conditions
below (\ref{s2val}). If we compare these restrictions with the Virasoro
conditions below (\ref{hvir}), we find that only the state
$h_{Vir}(p-2,0)$, or, equivalently, $h_{Vir}(0,q-2)$,
is missing at level 0. The conformal
weight of this missing state is $(p-2)(q-2)/4$.

The identification of the Virasoro minimal model by its primary operators
 therefore cannot be completed at level 0. For the case of the critical
 string and the Ising model, the authors of \cite{po2,we1,we2,hu1} indeed
 found one of the three primary operators of the Ising model,
 of weight $\tfrac{1}{2}$, at level 1.

At level 1 we will consider operators of the form
\begin{eqnarray}
\label{state}
   \Vop_1(p_2,s_1,s_2) &=& \gamma\, {\rm e}^{ip_2\phi_2+
             is_1\sigma_1+is_2\sigma_2}\,.
\end{eqnarray}
The condition $\Qop_1\Vop_1(p_2,s_1,s_2)=0$ determines the values of
 the momenta for which $\Vop_1$ is a physical state\footnote{The state
 $\Vop_1$ is the state of lowest ghost number at this level, and therefore
 cannot be written as $\Qop_1$ acting on any other level 1 state.}.
 This condition now comprises four quadratic equations in the momenta,
 of which the general solution is:
\begin{eqnarray}
  (A_1)&&p_2=-\tfrac{1}{2}\big(is_2 +i{\bch} -\sqrt{6-{\bch}^2}\big)\,,
 \nonumber\\
\label{stateA}
       &&s_1=0,\qquad s_2\ {\rm arbitrary}\,,
   \\
  (B_1)&&p_2=is_2 -\tfrac{1}{2}\big(i{\bch} +\sqrt{6-{\bch}^2}\big) \,,
 \nonumber\\
\label{stateB}
       &&s_1=\sqrt{3}\,s_2,\qquad s_2\ {\rm arbitrary}\,,
  \\
  (C_1)&&p_2=is_2 -\tfrac{1}{2}\big(i{\bch} +\sqrt{6-{\bch}^2}\big) \,,
 \nonumber\\
\label{stateC}
       &&s_1=-\sqrt{3}\big(s_2+\tfrac{2}{3}i\sqrt{6-{\bch}^2}\big),
          \qquad s_2\ {\rm arbitrary}\,,
\end{eqnarray}
As we did for level 0, we now choose ${\bch}$ equal to (\ref{Qmin}), and
 restrict the momenta $s_1,s_2$ to the allowed values (\ref{s1val},
 \ref{s2val}).
 However, at level 1 the momenta $s_1$ and $s_2$ are further
restricted by (\ref{stateA}-\ref{stateC}).
It is easy to see that these conditions, and the fact that
 $p$ and $q$ are relatively prime, imply:
\begin{eqnarray}
    s_1=0 &\rightarrow& r_2=t_2=0\,, \\
    s_1=\sqrt{3}s_2 &\rightarrow& r_1=t_1=0 \,, \\
    s_1=-\sqrt{3}\big(s_2+\tfrac{2}{3}i\sqrt{6-{\bch}^2}\big)
    &\rightarrow& {\rm no\ solution\ for}\ r_1,r_2,t_1,t_2 \,.
\end{eqnarray}
Therefore only the solutions (\ref{stateA}) and (\ref{stateB}) remain.
 On calculating the weight of these states we find the following
 Virasoro weights:
\begin{eqnarray}
\label{hvir1}
   (A_1) &\rightarrow& h_{Vir}(r_1,t_1+1)\,, \quad
   (B_1) \rightarrow   h_{Vir}(r_2+1,t_2) \,.
\end{eqnarray}
The state $h_{Vir}(p-2,0)$ is present in $(B_1)$, the state $h_{Vir}(0,q-2)$
 in $(A_1)$. Therefore the state which was missing at level is
 indeed found at level 1. Note that all the other Virasoro weights which
 were obtained at level 0 are again obtained at level 1, except the two
 states $h_{Vir}(0,0)$ and $h_{Vir}(p-2,q-2)$ of weight 0.

The weights found in solutions $(A_1)$ and $(B_1)$ take on the
 same values. This can be understood from applying twice the
 transformation (\ref{rt-trans1}), or, more directly, from the
 invariance of (\ref{hvir}) under the transformation
 $r\to p-2-r, t\to q-2-t$. The latter transformation gives a 1-1 map of
the Virasoro weights corresponding to the $(A_1)$ solution onto
those corresponding to the $(B_1)$ solution.

Thus all primary operators of the $(p,q)$ Virasoro minimal model are
 obtained at level 0 and 1 of the $\Qop_1$ cohomology.

Before we start to discuss the role of picture changing and screening
operators in all of this, we will first try to understand the cohomology
of $\Qop_1$ a bit further, in the spirit of \cite{li1b}. In this paper the
BRST cohomology of the noncritical string is computed, using a Felder
resolution of the minimal model in the matter sector, and using
a general result for the cohomology of the BRST operator when acting on
the tensor product of two Fock modules. The latter cohomology is quite
simple, for generic momenta the cohomology contains only the tachyonic
states, and for some special values of the momenta two extra states appear.
These are related to the existence of singular vectors in the Fock
modules. In the proof one replaces the matter and Liouville scalar fields
$\phi_M,\phi_L$ by linear combinations $\phi_M\pm i \phi_L$, which simplifies
the analysis considerably.

Let us now present some evidence that a similar structure also exists in the
case of the $\Qop_1$ cohomology acting on the tensor product of three
Fock spaces. First, we are going to perform a transformation of the fields
by defining
\begin{eqnarray}
\chi_1 & = & \partial(\phi_2-i\sigma_2) \,,\\
\chi_2 & = & \partial(\phi_2 +\tfrac{1}{2} i \sigma_2
   -\tfrac{1}{2} i \sigma_1 \sqrt{3})\,, \\
\chi_3 & = & \partial(\phi_2 +\tfrac{1}{2} i \sigma_2
      +\tfrac{1}{2} i \sigma_1 \sqrt{3}) \,.
\end{eqnarray}
The OPE's of the fields $\chi_i$ read
\begin{equation}
\chi_i(z) \chi_j(w) \sim \frac{-\frac{3}{2}(1-\delta_{ij}) }{(z-w)^2}\,.
\end{equation}
In addition, we introduce the parameter $\alpha$, defined by
\begin{eqnarray}
Q & = & \sqrt{\frac{3}{2}} (\alpha+\frac{1}{\alpha} ) \,,\\
\sqrt{6-Q^2} & = & i\sqrt{\frac{3}{2}} (\alpha-\frac{1}{\alpha})\,.
\end{eqnarray}
For the $(p,q)$ minimal $W_3$ model, the parameter $\alpha$ equals
$\sqrt{q/p}$.
In terms of these new variables and fields, the BRST current $j_1$ reads
\begin{eqnarray} j_1 & = & \frac{2i}{3} \sqrt{\frac{2}{3}}
\left( \gamma
\left\{ \chi_1(\chi_2\chi_3)-
\frac{1}{\alpha}\sqrt{\frac{3}{2}} \dif (\chi_1 \chi_2) -
\alpha\sqrt{\frac{3}{2}} \dif (\chi_2 \chi_3)
\right.   \right.   \nonumber\\
& & \qquad \qquad \,\,\, \left.  \left.
-\sqrt{\frac{3}{2}}
\left(\alpha (\chi_1 \dif \chi_3) +\frac{1}{\alpha}
(\dif \chi_1 \chi_3) \right) +
\left(\frac{3}{2\alpha^2}-\frac{3}{4}\right)\dif^2 \chi_1
\right.   \right.  \nonumber \\ & & \qquad \qquad \,\,\, \left.  \left.
 + \frac{3}{4} \dif^2 \chi_2 +
\left(\frac{3\alpha^2}{2} - \frac{3}{4}\right)\dif^2 \chi_3
\right\}
\right.  \nonumber   \\ & & \qquad \quad \left.
+ \frac{3}{2} \beta (\dif\gamma (\gamma (\chi_1+\chi_2+\chi_3)))
-\frac{3}{2} \sqrt{\frac{3}{2}} \left( \alpha+\frac{1}{\alpha}\right)
\dif\beta (\dif \gamma \gamma) \right) \,.
\end{eqnarray}

To continue along the lines of \cite{li1b}, we consider the mode
expansion of $\Qop_1$, acting on a tensor product of three
Fock modules and the ghost Hilbert space.
We can decompose it in terms of ghost zero modes as
\begin{equation}
\Qop_1=d-\beta_0 M + \gamma_0 W_0 + \gamma_0 \beta_0 H\,.
\end{equation}
Compared to the ordinary noncritical string, an extra complication
arises, in that a term containing $\gamma_0 \beta_0 $ appears.
The condition that $\Qop_1^2=0$ translates in the conditions
\begin{eqnarray}
d^2 & = & MW_0 \,,\\
M(d+H) & = & dM \,,\\
W_0 d & = & (d+H)W_0 \,,\\
(d+H)^2 & = &  W_0M \,.
\end{eqnarray}
The operator $d$ preserves the
subspace of states annihilated by $W_0$ and $\beta_0$, and on this space
it satisfies $d^2=0$, so that one can consider the cohomology of $d$ on
this subspace (this cohomology is usually
 called the relative cohomology of $\Qop_1$). In the case of the
non-critical Virasoro string, there is a simple relation between the
relative and absolute cohomology. Here that relation is less clear, since
$W_0$ is not diagonalizable in general.
We will ignore this problem
for the moment and focus only on $d$. It can be written as the sum
of a piece containing the terms linear in the oscillators and the ghosts and
a set of remaining terms. We will be only interested in the piece linear in
the ghosts and the oscillators. Denoting the modes of $\chi_i$ by $\eta^i_n$,
so that $\chi_i=\sum \eta^i_n z^{-n-1}$, we find for this part of
$d$
\ba \label{eqq1} &
-\frac{2i}{3}\sqrt\frac{2}{3}
\sum_{n\neq 0} \gamma_{-n} ( \eta^1_n P_2^+(n)P_3(n) +
\eta^2_n P_1(n) P_3(n) + \eta^3_n P_1(n) P_2^-(n) ) & \nonumber
\\ &
-\frac{i}{\sqrt{6}}
\sum_{n \neq 0} \gamma_{-n} (n^2+n)(\eta^1_n+\eta^2_n+\eta^3_n)\,, &
\ea
where the polynomials $P_1,P_2^{\pm}$ and $P_3$ are given by
\begin{eqnarray}
P_1(n) & = & p_2-is_2 +i\sqrt{\frac{3}{2}}\alpha (n+2)\,, \\
P_2^+(n) & = & p_2 +\frac{i}{2} s_2 -\frac{i}{2} s_1 \sqrt{3}+
i\sqrt{\frac{3}{2}} \left( \frac{n+1}{\alpha} +\alpha\right) \,,\\
P_2^-(n) & = & p_2 +\frac{i}{2} s_2 -\frac{i}{2} s_1 \sqrt{3}+
i\sqrt{\frac{3}{2}} \left(\alpha(n+1)+\frac{1}{\alpha} \right) \,,\\
P_3(n) & = & p_2 +\frac{i}{2} s_2 +\frac{i}{2} s_1 \sqrt{3}+
i\sqrt{\frac{3}{2}} \frac{n+2}{\alpha} \,.
\end{eqnarray}
The almost factorized form of $d$ in (\ref{eqq1}) is very suggestive.
Maybe one can define a modified operator $d$ in such a way that the last term
in (\ref{eqq1}) is absent.
In any case numerical evidence shows that all the states in the cohomology of
$\Qop_1$ have values of the momenta related to zeros of the polynomials $P_i$.
Restricting attention to Virasoro primaries, we conjecture that the
relative cohomology of $\Qop_1$ can be organized as follows
\begin{itemize}
\item[-] For generic momenta there is only one state at ghost number two and
level zero, the standard tachyonic state, if $P_1(0)P_2(0)P_3(0)=0$.
(Here $P_2(0)\equiv P_2^+(0)=P_2^-(0))$.
\item[-] If integers $u_1,u_2>0$ exist such that (i) $P_1(u_1)=P_2^+(u_2)=0$,
(ii) $P_2^-(u_1)=P_3(u_2)=0$, or (iii) $P_1(u_1)=P_3(u_2)=0$ there are two
states at level $u_1 u_2$ and ghost numbers $2,3$.
\item[-] If integers $u_1,u_2<0$ exist such that (i) $P_1(u_1)=P_2^+(u_2)=0$,
(ii) $P_2^-(u_1)=P_3(u_2)=0$, or (iii) $P_1(u_1)=P_3(u_2)=0$ there are two
states at level $u_1 u_2$ and ghost numbers $1,2$.
\item[-] If integers $u_1,u_2,u_3,u_4$ exist such that $P_1(u_1)=P_3(u_4)=0$,
and $P_2^-(u_2)+u_3\alpha=0$, $P_2^+(u_3)+u_2/\alpha=0$, there may be extra
states at level $u_1u_3-u_3u_2+u_2u_4$ and at three consecutive ghost numbers.
The precise ghost numbers depend on the signs of the $u_i$ but one of them
is 2.
\end{itemize}
The three kinds of states are related to the six kinds of null states that
minimal $W_3$ representations generically have. These six are reduced to three
here since we did not yet take the $\Qop_0$ cohomology. This is all
similar very
to the analysis for the noncritical ordinary string, where a similar though
simpler result holds for the cohomology of the BRST operator on the
product of two Fock spaces and the ghost Hilbert space.

To prove this conjecture, and specify more precisely the cohomology for the
last of the four cases, one could try to use the same spectral sequence
techniques as in \cite{li1b}. In this respect it is suggestive that
in terms of the $\chi_i$ no term in $\Qop_1$ contains any $\chi_i$ more
than once, and if one would assign a positive degree to say $\chi_1$ and
negative degree to the others, the first term in the spectral sequence
would be the cohomology of the $\sum \gamma_{-n} \eta^1_n$
term in (\ref{eqq1}).
Apart from the problem with the extra term in (\ref{eqq1}), this would
show that one needs zeros of the polynomials $P_i$ to get extra states,
but the details are not clear to us.

As an example, the third case with $u_1=u_2=-1$ yields precisely the series
of states (\ref{stateA}), (\ref{stateB}) and (\ref{stateC}).

\vspace{0.5truecm}
\noindent{\bf 4.\ Screening and picture changing operators}
\vspace{0.5truecm}

In the previous section we discussed in detail
the $\Qop_1$ cohomology for level
 0 and 1, and for the particular ghost structures (\ref{vac}, \ref{state}).
In addition, we conjectured for which momenta there will be extra states at
other levels.

In the case of the critical $W_3$ string it has been proposed that all
these states can be obtained from the states at level 0 and 1 by acting on
them with picture changing and screening operators \cite{po2,we1,we2,hu1}.
Generalizing the case of the critical string, we
 conjecture that
 with ${\bch}$ equal to $Q_{min}$ (\ref{Qmin}) and the
 momenta of the Liouville fields
 restricted to those of the $(p,q)$ minimal model,
 all states in the $\Qop_1$ cohomology will be related by picture
 changing and screening operators to the states obtained thus far, or
to descendants of the primary operators discovered at level 0 and 1.

Screening operators (anti-)commute with $\Qop_1$, and have weight 0.
 They transform physical states into other physical states, provided
 the action of such an operator on a state is well-defined (see \cite{we2},
 and the discussion below). Screening operators may change the level.

Picture changing operators are of the form $[\Qop,\phi]$, where
 $\phi$ is a scalar field.
 They also produce solutions of the cohomology when acting on a
 physical state. In our case there are
 three such operators, corresponding to $\phi_2,\sigma_1$ and
 $\sigma_2$. They have weight 0, change the
 ghost number of a physical state by a single unit but don't change
the level.  Applying the same picture changing
 operator twice gives 0.

We will not prove the above conjecture in this paper. We will
 provide evidence for it by showing that the weights of the
 Virasoro $(p,q)$ minimal model occur exactly once among the
 level 0 and 1 states which we obtained in the previous section,
 if states related by screening and picture changing operators
 are identified.  This illustrates the
 use of these operators, and hopefully will stimulate efforts in
 obtaining a proof.

Let us first give a list of the screening operators
 which are relevant for
 our calculation. There are two screening operators
 of the form
\begin{eqnarray}
\label{Si}
   S_i &=& \oint dz \,\beta(z) {\rm e}^{i\alpha_{2,S_i} \phi_2(z)
     +i\beta_{1,S_i}\sigma_1(z) + i\beta_{2,S_i}\sigma_2(z)} \,.
\end{eqnarray}
These operators anticommute with $\Qop_1$ provided the momenta $\alpha_2,
 \beta_1,\beta_2$ take on the following values:
\begin{eqnarray}
  S_1\ : \ &&\alpha_{2,S_1} = \tfrac{1}{3}\{i{\bch} + \sqrt{6-{\bch}^2}\} \,,
  \nonumber\\
           &&\beta_{1,S_1} =
    \tfrac{1}{6}i\sqrt{3}\{i{\bch} + \sqrt{6-{\bch}^2}\} \,,
  \nonumber\\
        \label{S1}
           &&\beta_{2,S_1} =
    \tfrac{1}{6}i\{i{\bch} + \sqrt{6-{\bch}^2}\} \,, \\
  S_2\ : \ &&\alpha_{2,S_2} = \tfrac{1}{3}\{i{\bch} - \sqrt{6-{\bch}^2}\} \,,
  \nonumber\\
           &&\beta_{1,S_2}=0 \,,
  \nonumber\\
        \label{S2}
           &&\beta_{2,S_2} = -\tfrac{1}{3}i\{i{\bch} - \sqrt{6-{\bch}^2}\} \,.
\end{eqnarray}
Then there are four screening operators which involve only the Liouville
 fields, and which are of the form
\begin{eqnarray}
\label{Ti}
     T_i &=& \oint dz\, {\rm e}^{i\beta_{1,T_i}\sigma_1(z) +
                      i\beta_{2,T_i}\sigma_2(z)} \,.
\end{eqnarray}
They commute with $\Qop_1$ for the following values of the momenta:
\begin{eqnarray}
 T_1\ : \ &&\beta_{1,T_1} =
 \big({\bch} - i\sqrt{6-{\bch}^2} \big)/\sqrt{3} \,,
   \nonumber\\
 \label{T1}
            &&\beta_{2,T_1} = 0\,,  \\
 T_2\ : \ &&\beta_{1,T_2} =
  -\big({\bch} + i\sqrt{6-{\bch}^2} \big)/\sqrt{3} \,,
   \nonumber\\
 \label{T2}
            &&\beta_{2,T_2} = 0\,,   \\
 T_3\ : \ &&\beta_{1,T_3} =
   \tfrac{1}{2}\big(-{\bch} + i\sqrt{6-{\bch}^2} \big)/\sqrt{3} \,,
   \nonumber\\
 \label{T3}
            &&\beta_{2,T_3} =
   -\tfrac{1}{2}\big(-{\bch} + i\sqrt{6-{\bch}^2} \big) \,,  \\
 T_4\ : \ &&\beta_{1,T_4} =
   \tfrac{1}{2}\big({\bch} + i\sqrt{6-{\bch}^2} \big)/\sqrt{3} \,,
   \nonumber\\
 \label{T4}
            &&\beta_{2,T_4} =
   -\tfrac{1}{2}\big({\bch} + i\sqrt{6-{\bch}^2} \big) \,.
\end{eqnarray}
Finally, we have two screening operators of the form:
\begin{eqnarray}
\label{Ri}
   R_i &=& \oint dz \,\gamma(z) {\rm e}^{i\alpha_{2,R_i} \phi_2(z)}
          \,,
\end{eqnarray}
with the corresponding momenta:
\begin{eqnarray}
\label{R1}
  R_1\ : \ &&\alpha_{2,R_1} = -i{\bch} - \sqrt{6-{\bch}^2} \,, \\
\label{R2}
  R_2\ : \ &&\alpha_{2,R_2} =  -i{\bch} + \sqrt{6-{\bch}^2} \,.
\end{eqnarray}

The three picture changing operators are $P_{\phi_2}=[\Qop_1,\phi_2]$,
 $P_{\sigma_1}=[\Qop_1,\sigma_1]$ and
 $P_{\sigma_2}=[\Qop_1,\sigma_2]$. We find for $P_{\phi_2}$:
\begin{eqnarray}
  P_{\phi_2}&=& -{i\over 3\sqrt{6}}\bigg[
     12\gamma(\p)^2 + 12{\bch}(\partial\gamma)\p +
     (-15+4{\bch}^2)\partial^2\gamma\bigg] \nonumber\\
\label{P2}
     && + {2i\over \sqrt{6}}\gamma T_L
     + i\sqrt{6}\gamma(\partial\gamma)\beta\,,
\end{eqnarray}
and similar expressions for $P_{\sigma_1}$ and $P_{\sigma_2}$. In this
 section we consider only the action of these picture changing operators
 on the vacuum state (\ref{vac}). The only term which then contributes
 is the $\partial^2\gamma$-contribution, which is
 present in all three picture changing operators. It produces
 physical states
\begin{eqnarray}
\label{vac1}
  \bar\Vop_0(p_2,s_1,s_2) &=&
  (\partial^2\gamma)(\partial{\gamma})\gamma\, {\rm e}^{ip_2\phi_2+
             is_1\sigma_1+is_2\sigma_2} \,,
\end{eqnarray}
for the same values of the momenta obtained in (\ref{vacA}-\ref{vacC}).
 Since all picture changing operators act
 the same way in this case, we will denote them collectively by $P$.

The action of the screening operators is more complicated. All our
 screening operators are of the form
\begin{eqnarray}
 &&S_i=\oint dz_i K_i(\beta(z_i),\gamma(z_i))\,
 {\rm exp}(i\sum_m p_{m,S_i} \phi_m(z_i)) \,,
\end{eqnarray}
 where $\phi_m$ is a set of scalar
 fields, and $p_{m,S_i}$ are the screening momenta in the operator
 $S_i$ for the $m$'th field. These operators act on states of the form
\begin{eqnarray}
 {\bf O}=L(\beta(w),\gamma(w))\, {\rm exp}(i\sum_m p_{m} \phi_m(w)) \,.
\end{eqnarray}
 The condition
 under which this action is well-defined is discussed in detail
 in \cite{we2}. If the action of
 the product of $n$ such screening operators on ${\bf O}$ is considered,
 then the number
\begin{eqnarray}
\label{Pn}
   P_n =  n-1 + \sum_m \sum_{i,j=1,\ i<j}^n p_{m,S_i}p_{m,S_j}
          + \sum_m \sum_{i=1}^n p_{m,S_i} p_m
\end{eqnarray}
should be an integer. This condition arises from the fact that the
 succesive OPE's give, after appropriate changes in
 the integration variables $z_i$, rise to a single factor $(z_1-w)^{P_n}$.
 The OPE's of the ghost contributions will similarly give a factor
 $(z_1-w)^{P_{gh}}$, where $P_{gh}$ is guaranteed to be an integer.
 The integral over $z_1$  then gives a well-defined and non-trivial
 result if
\begin{eqnarray}
\label{Pcond}
   P_n+P_{gh}=-1\,.
\end{eqnarray}

The momenta of the final state,
 ${\bf O}'=S_1\ldots S_n {\bf O}$, if it is defined, are equal
 to $p_m + \sum_i p_{m,S_i}$, and therefore the conformal weight
 of this proposed
 ${\bf O}'$ can be calculated. Using the condition that
 the weight of the screening operators is zero, and,
 independently, that
 the weights of ${\bf O'}$ and ${\bf O}$ are equal, it is
 not difficult to show that
\begin{eqnarray}
\label{Pnn}
   P_n = h_{L,{\bf O}} - h_{L,{\bf O'}}  -1 + \sum_i h_{K_i},
\end{eqnarray}
where $h_{K_i},\ h_{L,{\bf O}}, h_{L,{\bf O'}}$ are the
 conformal weights of the ghost contributions to $S_i,\ {\bf O}$ and
 ${\bf O'}$ respectively. Therefore, if the conformal weights of the
 initial
 and of the proposed final state are equal, and if the
 momenta of the screening operators interpolate between the
 momenta of initial and final states, $P_n$ is automatically
 integer and can be easily calculated.

In discussing the action of the screening operators, it is useful to
 characterize their effect on the momenta, and on the labels
 $r_1,r_2,t_1,t_2$ in the Liouville sector. Consider a screening
 operator
\begin{eqnarray}
\label{Sgen}
   S&=& (R_1)^{l_1}(R_2)^{l_2}(S_1)^{m_1}(S_2)^{m_2}(T_1)^{n_1}(T_2)^{n_2}
        (T_3)^{n_3}(T_4)^{n_4}\,.
\end{eqnarray}
If we choose a $(p,q)$ $W_3$ minimal model, so that $Q=Q_{min}$
with $Q_{min}$ given in (\ref{Qmin}),
the changes in the momenta due to (\ref{Sgen})
 are given by:
\begin{eqnarray}
\label{dp2}
   \Delta p_2&=& i\big((2m_1-6l_1)q+(2m_2-6l_2)p\big)/\sqrt{6pq}\,,\\
\label{ds1}
   \Delta s_1&=& \sqrt{3}\big( (-m_1+2n_1-n_3)q+(-2n_2+n_4)p\big)/
     \sqrt{6pq} \,,\\
\label{ds2}
   \Delta s_2&=& \big( (-m_1+3n_3)q+(2m_2-3n_4)p\big)/\sqrt{6pq} \,.
\end{eqnarray}
The change in the Liouville momenta induces changes in the labels
 $r_1,r_2,t_1,t_2$ in (\ref{s1val}, \ref{s2val}). These are
 given by
\begin{eqnarray}
   \Delta r_1&=& n_1-2n_3+lp\,,\nonumber\\
   \Delta r_2&=&m_1-2n_1+n_3+kp\,,\nonumber\\
   \Delta t_1&=&m_2+n_2-2n_4+lq\,,\nonumber\\
\label{drt}
   \Delta t_2&=&-2n_2+n_4+kq\,,
\end{eqnarray}
where $l$ and $k$ are integers chosen in such a way that
 the $\Delta r_i$ and $\Delta t_i$ produces labels in the allowed
 range $0\le r_1+r_2\le p-3,\ 0\le t_1+t_2\le q-3$. Of course, the
 action of $S$ will be well-defined only if the initial and final
 conformal weights are equal.

Now that the action of the screening operators has been clarified,
 let us first discuss the way they act on the states at level 0. We
 choose a $(p,q)$ $W_3$ minimal model. Then the Virasoro weights
 for, e.g., the physical states $(A_0)$, as given in (\ref{hvir0}),
 occur with a certain multiplicity. These weights are given
 by $h_{Vir}(r_2,t_2)$, and thus independent of $r_1$ and $t_1$.
 Using (\ref{drt}) it is easy to see that one can pass between the
 states of different $r_1$ and $t_1$ with screening operators. We
 have on the states $(A_0)$:
\begin{eqnarray}
{\rm On}\ (A_0)\quad
  && S_2P\quad {\rm gives}\quad \Delta\,(r_1,r_2,t_1,t_2) = (0,0,1,0)\,,
     \nonumber\\
 \label{ScrvacA}
  && S_1T_1T_3P\quad {\rm gives}\quad
        \Delta\,(r_1,r_2,t_1,t_2) = (-1,0,0,0)\,.
\end{eqnarray}
The presence of the picture changing operator $P$ is required
 because the extra $\partial^2\gamma$ it introduces in $\Vop_0$
 ensures that in the OPE with the screening operators (\ref{Pcond})
 is satisfied.

In the previous section we showed that the physical
 states $(A_0),(B_0)$ and $(C_0)$ are related by the discrete
 transformation (\ref{rt-trans1}). This induces transformations
 similar to (\ref{ScrvacA}) for the states $(B_0)$ and $(C_0)$.
 These can again be read off from (\ref{drt}) and are given by
 \begin{eqnarray}
{\rm On}\ (B_0)\quad
  && S_2T_4P\quad {\rm gives}\quad \Delta\,(r_1,r_2,t_1,t_2) = (0,0,-1,1)\,,
     \nonumber\\
 \label{ScrvacB}
  && S_1T_1P\quad
   {\rm gives}\quad \Delta\,(r_1,r_2,t_1,t_2) = (1,-1,0,0)\,,\\
{\rm On}\ (C_0)\quad
  && S_2T_2T_4P\quad {\rm gives}\quad
     \Delta\,(r_1,r_2,t_1,t_2) = (0,0,0,-1)\,,
     \nonumber\\
 \label{ScrvacC}
  && S_1P\quad {\rm gives}\quad \Delta\,(r_1,r_2,t_1,t_2) = (0,1,0,0)\,.
\end{eqnarray}

In fact, the relation between (\ref{vacA}-\ref{vacC}) due to the
 discrete symmetry (\ref{rt-trans1}) can be represented by
 screening operators. Consider again the solutions $(A_0)$, with
 Virasoro weights $h_{Vir}(r_2,t_2)$. If we perform a
 transformation (\ref{rt-trans1}) on the labels $(r_i,t_i)$, we
 find that $1+r_1'+r_2'=p-2-r_2,\ 1+t_1'+t_2'=q-2-t_2$. This means
 that $h_{Vir}(1+r_1'+r_2',1+t_1'+t_2')=
 h_{Vir}(p-2-r_2,q-2-t_2)=h_{Vir}(r_2,t_2)$. Therefore, with this
 transformation on the labels we find solutions among the $(B_0)$
 states
 with the same weight as the $(A_0)$ state.
 Given this change in the labels we use
 (\ref{drt}) to obtain the
 corresponding screening operator. We find that from any physical
 state $(A_0)$
 a state of type $(B_0)$ can be obtained using
\begin{eqnarray}
  && \Vop_{0,(B_0)}(p-3-r_1-r_2,r_1,q-3-t_1-t_2,t_1) =  \nonumber\\
\label{ScrvacBA}
  &&\qquad(T_1)^{1+r_2}
   (T_2)^{1+t_2}(T_3)^{2+r_1+r_2}(T_4)^{2+t_1+t_2}\,
   \Vop_{0,(A_0)}(r_1,r_2,t_1,t_2)\,.
\end{eqnarray}
In a similar way one obtains a relation between the states $(C_0)$ and
 $(A_0)$:
\begin{eqnarray}
  && \Vop_{0,(C_0)}(r_2,p-3-r_1-r_2,t_2,q-3-t_1-t_2) = \nonumber\\
\label{ScrvacCA}
  &&\qquad (T_1)^{2+r_1+r_2}
   (T_2)^{2+t_1+t_2}(T_3)^{1+r_1}(T_4)^{1+t_1}\,
   \Vop_{0,(A_0)}(r_1,r_2,t_1,t_2)\,.
\end{eqnarray}

If we set up an equivalence relation between states, under which two
 states that are related by screening operators are equivalent, then
 at this stage we can limit ourselves at level 0 to the states
 $(A_0)$, with the further restriction that from states which
 differ only in the labels $(r_1,t_1)$ only one representative
 is considered. The restrictions on $(r_2,t_2)$ are then given by
 $0\le r_2\le p-3,\ 0\le t_2\le q-3$. The
 multiplicity of the remaining
 Virasoro weights can still be equal to two.
 The weights with $1\le r_2\le p-3,\ 1\le t_2\le q-3$ all occur twice,
 since both the labels $(r_2,t_2)$ and $(p-2-r,q-2-t)$ are in the
 allowed range. The labels with either $r_2=0$ or $t_2=0$ occur only
 once. This doubling occurs for $p\ge 5,\ q\ge 4$.

This remaining doubling of the Virasoro weights can also be lifted by
 screening operators. We find for all
 $(p,q), p\ge 5,\ q\ge 4$:
\begin{eqnarray}
  &&\Vop_{0,(A_0)}(0,p-2-r_2,0,q-2-r_2) = \nonumber\\
  &&\qquad
   R_1(S_2)^2(T_1)^{r_2}(T_2)^{t_2+1}(T_3)^{p-2}(T_4)^qP\times
 \nonumber\\
\label{ScrvacAA}
 &&\qquad\qquad
  \times\Vop_{0,(A_0)}(p-4-r_2,r_2,q-3-t_2,t_2) \,.
\end{eqnarray}
This relates one of the states with $(r_2,t_2)$ to one of the
 states with $(p-2-r_2,q-2-r_2)$. Up to equivalence by
 screening operators, each Virasoro weight (except $(p-2)(q-2)/4$)
 therefore occurs
 once and only once at level 0.

In this application of the screening operators at level 0 we choose
 the screening operators such that they give the
 required $\Delta\,(r_1,r_2,t_1,t_2)$. By construction, the
 screening operators interpolate between the initial and final
 momenta, and leave the weight invariant. Therefore, as explained
 above, the factor $P_n$ (\ref{Pn}) is an integer.
 In the application (\ref{ScrvacA}-\ref{ScrvacC}) $P_n$ is equal to
 $2$, while the OPE of the ghost $\beta$
 in $S_i$ with $(\partial^2\gamma)(\partial\gamma)\gamma$
 gives $P_{gh}=-3$ (as well as
 less singular terms). In the second application (\ref{ScrvacBA},
 \ref{ScrvacCA}) we find immediately $P_n=-1$. In the third
 application (\ref{ScrvacAA}) we have an additional $\gamma$ from
 $R_1$. Therefore $P_n$ equals 3. The OPE of the ghosts
 gives $P_{gh}=-4$, leaving again the combination
 $(\partial\gamma)\gamma$. Therefore in all cases (\ref{Pcond}) is
 satisfied.

Now consider the states at level 1. Here we want to show that all states
 except those with momentum $(p-2)(q-2)/4$ can be obtained from the
 level 0 states using screening operators. Since the momenta of all
 level 0 and level 1 states are given in the previous section, it
 is a simple matter to use again (\ref{dp2}-\ref{drt}) to construct the
 appropriate screening operators. A useful hint about this choice follows
 from the values of the Virasoro weights. For instance, the states $(A_1)$
 have $r_2=t_2=0$, and weight $h_{Vir}(r_1,t_1+1)$. For the level 0 states
 with $r_2,t_2=0$ the weights are
 $h_{Vir}(0,0)=0,h_{Vir}(1+r_1,1+t_1),h_{Vir}(r_1,t_1)$
 for solutions $(A_0),(B_0),(C_0)$, respectively. Since
 $h_{Vir}=0$ does not occur at level 1 (see the discussion at the end
 of Section 3), the states $(A_1)$ cannot be obtained from
 states $(A_0)$ with $r_2=t_2=0$. They
 can
 be produced from $(B_0)$ and/or $(C_0)$. For $(B_0)$ we must
 choose $\Delta r_1=1,\Delta t_1=0$, for $(C_0)$
 $\Delta r_1=0,\Delta t_1=-1$, with of course $\Delta r_2=\Delta t_2=0$.

This leads to the following result. The states $(A_1)$ can be obtained
 from level 0 by the following operators
\begin{eqnarray}
  \Vop_{1,(A_1)}(r_1,0,t_1,0) &=& (S_2)^2T_2(T_4)^2P\,
        \Vop_{0, (C_0)} (r_1,0,t_1+1,0)
 \nonumber\\
\label{ScrA1}
      &=& (S_1)^2T_1P\, \Vop_{0, (B_0)} (r_1-1,0,t_1,0) \,.
\end{eqnarray}
Depending on the values of the labels on the right-hand-side, both or
 only one of the above transitions is allowed.
 The picture changing operator is again required to ensure
 that (\ref{Pcond}) is satisfied. Note that the state
 $\Vop_{1,(A_1)}(0,0,q-3,0)$,
 corresponding to the missing primary operator at
 level 0, cannot be obtained in this way because then the values of
 the labels on the right-hand-side of (\ref{ScrA1}) fall outside the
 allowed range.

Similarly, for the solutions $B_1$ at level 1 we find that
\begin{eqnarray}
  \Vop_{1,(B_1)}(0,r_2,0,t_2)
     &=& (S_1)^2(T_1)^2T_3P\, \Vop_{0, (A_0)} (0,r_2+1,0,t_2)
 \nonumber\\
\label{ScrB1}
            &=& (S_2)^2T_4P \,\Vop_{0, (B_0)} (0,r_2,0,t_2-1) \,.
\end{eqnarray}

As a last point, we must now show that a screening operator connects
 the two states with Virasoro weight $(p-2)(q-2)/4$, which does not
 occur at level 0. In fact, we can connect every state
 $\Vop_{1,(B_1)}$ to a state $\Vop_{1,(A_1)}$ by screening operators,
 analogously the the relations (\ref{ScrvacBA}-\ref{ScrvacCA}) at level
 0. A look at the form of the Virasoro weights (\ref{hvir1}) tells us
 that we should  correlate the labels $(r_1,0,t_1,0)$ for $(A_1)$
 and  $(0,r_2=p-3-r_1,0,t_2=q-3-t_1)$ at $(B_1)$. This leaves the
 the $\phi_2$-momentum invariant, and corresponds to a transformation
 (\ref{rt-trans1}) in the Liouville-sector.
 It can be easily seen
 that this correspondence is realized by:
\begin{eqnarray}
  &&\Vop_{1,(B_1)}(0,p-3-r_1,0,q-3-t_1) = \nonumber\\
\label{Scr1BA}
   &&\qquad\qquad (T_1)^{2+r_1}
     (T_2)^{2+t_1}(T_3)^{1+r_1}(T_4)^{1+t_1}
       \Vop_{1,(A_1)}(r_1,0,t_1,0) \,.
\end{eqnarray}
Thus all primary operators of the $(p,q)$ Virasoro minimal model
 occur with multiplicity 1, up to screening and picture changing
 operators.

\vspace{0.5truecm}
\noindent{\bf 5.\ Examples}
\vspace{0.5truecm}

In this section we will illustrate the results from the previous
 sections for the cases $(p,q)=(4,3)$ and $(5,4)$.

In the first case $c_3^{3;(4,3)}=1/2$, so that the Virasoro minimal
 model corresponds to the Ising model. The value
 of the background charge in this case is:
\begin{eqnarray}
  (p,q)=(4,3)\to {\bch}= 21/\sqrt{72},\quad
  \sqrt{6-{\bch}^2}= - 3i/\sqrt{72} \,.
\end{eqnarray}
This case has been much
 studied recently in the case of the critical $W_3$-string
 \cite{po2,we1,we2,hu1}. In our
 construction, the $W_3$ minimal model we start out with
 has $c_l^{(4,3)}=0$, so that the Liouville sector corresponds to
 the ``trivial'' $W_3$ minimal model.

In Table 2 we present the momenta and conformal weights for the
 physical states (\ref{vacA}). As we saw in the previous section,
 the choices (\ref{vacB}-\ref{vacC}) are connected to (\ref{vacA})
 by screening operators, and in this sense equivalent.
The three possible choices of the labels $r_1,r_2,t_1,t_2$
are related to each other by the discrete transformation (\ref{rt-trans1}),
so that the Liouville weight $h_l$ is equal for the three states. There
is only a three-fold degeneracy, because under (\ref{rt-trans2}) no new
labels are generated. The Virasoro weight $h_{Vir}$ for these states is
completely determined by $(r_2,t_2)$ (\ref{hvir0}), so that the equality
of $h_{Vir}$ for $(0,0,0,0)$ and $(1,0,0,0)$ is understood. In the
previous section we showed that these two states can also be related
by screening and picture changing operators. Note that the $\phi_2$-momenta
for these two states are conjugate to each other, where conjugation of
a momentum $p_\phi$ for a field with background charge ${\bch}_\phi$
is defined as
\begin{eqnarray}
\label{conj}
   (p_\phi)^* \equiv -p_\phi - 2i{\bch}_\phi \,.
\end{eqnarray}
However, the Liouville momenta are not related by conjugation, since the
 lattice chosen in (\ref{s1val}-\ref{s2val}) does not transform into
 itself under conjugation.

At level 0 we therefore find that the two available
 Virasoro weights occur with multiplicity 1, so that the
 operations (\ref{ScrvacAA}) are not required.

The level 1 states of type $\Vop_{1,(A_1)}$ are given in Table 3. Here
 there are only two possible states. We showed in the previous section
 that the state $(1,0,0,0)$ is related to $\Vop_{0,(B_0)}$
 (\ref{ScrA1}), and, by implication, therefore also to
 $\Vop_{0,(A_0)}$.

\vspace{.25cm}
\begin{center}
\renewcommand{\arraystretch}{1.5}
\begin{tabular}{|l||l|l|l|l|l|}
\hline
$(r_1,r_2,t_1,t_2)$ &\hfil $p_2\sqrt{72}$\hfil&\hfil $s_1\sqrt{72}$\hfil
  &\hfil $s_2\sqrt{72}$\hfil
 &\hfil $h_l$\hfil&\hfil $h_{Vir}$\hfil\\
\hline
\hfil$(0,0,0,0)$\hfil&\hfil$-18i$\hfil&
     \hfil $0$\hfil&\hfil $0$\hfil&
          \hfil $0$\hfil &\hfil $0$\hfil\\
\hfil$(0,1,0,0)$\hfil&\hfil$-21i$\hfil&
     \hfil$-3\sqrt{3}$\hfil&\hfil $-3$\hfil&
          \hfil $0$\hfil & \hfil $1/16$\hfil\\
\hfil$(1,0,0,0)$\hfil&\hfil$-24i$\hfil&
     \hfil $0$\hfil&\hfil $-6$\hfil&
          \hfil $0$\hfil & \hfil $0$\hfil\\
\hline
\end{tabular}
\renewcommand{\arraystretch}{1.0}
\end{center}
\vspace{.25truecm}

\noindent {\bf Table\ 2.} \ \ \ \ \ Momenta and conformal weights for
 the states $\Vop_{0,(A_0)}$ (\ref{vacA})
 for $p=4,\ q=3$. Note that all momenta in the table have been
 multiplied by a factor $\sqrt{72}$. $h_l$ is the
 contribution of the Liouville sector to the total conformal
 weight $h_{Vir}$.
\vspace{.5truecm}

\vspace{.25cm}
\begin{center}
\renewcommand{\arraystretch}{1.5}
\begin{tabular}{|l||l|l|l|l|l|}
\hline
$(r_1,r_2,t_1,t_2)$ &\hfil $p_2\sqrt{72}$\hfil&\hfil $s_1\sqrt{72}$\hfil
  &\hfil $s_2\sqrt{72}$\hfil
 &\hfil $h_l$\hfil&\hfil $h_{Vir}$\hfil\\
\hline
\hfil$(0,0,0,0)$\hfil&\hfil$-12i$\hfil&
     \hfil $0$\hfil&\hfil $0$\hfil&
          \hfil $0$\hfil &\hfil $1/2$\hfil\\
\hfil$(1,0,0,0)$\hfil&\hfil$-9i$\hfil&
     \hfil $0$\hfil&\hfil $-6$\hfil&
          \hfil $0$\hfil & \hfil $1/16$\hfil\\
\hline
\end{tabular}
\renewcommand{\arraystretch}{1.0}
\end{center}
\vspace{.25truecm}
\noindent {\bf Table\ 3.} \ \ \ \ \ Momenta and conformal weights for
 the states $\Vop_{1,(A_1)}$ (\ref{stateA})
 for $p=4,\ q=3$.
\vspace{.5truecm}

As we have seen, the $(4,3)$ case does not contain all the features
 discussed in the previous sections. The case $(p,q)=(5,4)$ corresponds
 more closely to the generic situation. In this cases we have
\begin{eqnarray}
  (p,q)=(5,4)\to {\bch}= 27/\sqrt{120},\quad
  \sqrt{6-{\bch}^2}= - 3i/\sqrt{120} \,.
\end{eqnarray}
The Liouville sector corresponds to a $W_3$ minimal model with
 $c_l^{(5,4)}=4/5$, and the $\Qop_1$ cohomology will result
 in the states of the $c_3^{3;(5,4)}=7/10$ Virasoro minimal model.

The number of physical states $(A_0)$ at level 0 is equal to
\begin{eqnarray}
   \left({p-1 \atop 2}\right)\left( {q-1 \atop 2} \right)
\end{eqnarray}
and therefore increases quadratically with $p$ and $q$. The 18 states
$(A_0)$ for $p=5,\ q=4$ are presented in Table 4. The multiplicity
of the Liouville weights is either 3 or 6, as explained in Section 3.

\vspace{.25cm}
\begin{center}
\renewcommand{\arraystretch}{1.5}
\begin{tabular}{|l||l|l|l|l|l|}
\hline
$(r_1,r_2,t_1,t_2)$ &\hfil $p_2\sqrt{120}$\hfil&\hfil $s_1\sqrt{120}$\hfil
  &\hfil $s_2\sqrt{120}$\hfil
 &\hfil $h_l$\hfil&\hfil $h_{Vir}$\hfil\\
\hline
\hfil$(0,0,0,0)$\hfil&\hfil$-24i$\hfil&
     \hfil $0$\hfil&\hfil $0$\hfil&
          \hfil $0$\hfil &\hfil $0$\hfil\\
\hfil$(0,0,0,1)$\hfil&\hfil$-19i$\hfil&
     \hfil$5\sqrt{3}$\hfil&\hfil $5$\hfil&
          \hfil $2/3$\hfil & \hfil $7/16$\hfil\\
\hfil$(0,0,1,0)$\hfil&\hfil$-14i$\hfil&
     \hfil $0$\hfil&\hfil $10$\hfil&
          \hfil $2/3$\hfil & \hfil $0$\hfil\\
\hfil$(0,1,0,0)$\hfil&\hfil$-28i$\hfil&
     \hfil $-4\sqrt{3}$\hfil&\hfil $-4$\hfil&
          \hfil $1/15$\hfil &\hfil $1/10$\hfil\\
\hfil$(0,1,0,1)$\hfil&\hfil$-23i$\hfil&
     \hfil $\sqrt{3}$\hfil&\hfil $1$\hfil&
          \hfil $1/15$\hfil &\hfil $3/80$\hfil\\
\hfil$(0,1,1,0)$\hfil&\hfil$-18i$\hfil&
     \hfil $-4\sqrt{3}$\hfil&\hfil $6$\hfil&
          \hfil $2/5$\hfil &\hfil $1/10$\hfil\\
\hfil$(1,0,0,0)$\hfil&\hfil$-32i$\hfil&
     \hfil $0$\hfil&\hfil $-8$\hfil&
          \hfil $1/15$\hfil &\hfil $0$\hfil\\
\hfil$(1,0,0,1)$\hfil&\hfil$-27i$\hfil&
     \hfil $5\sqrt{3}$\hfil&\hfil $-3$\hfil&
          \hfil $2/5$\hfil &\hfil $7/16$\hfil\\
\hfil$(1,0,1,0)$\hfil&\hfil$-22i$\hfil&
     \hfil $0$\hfil&\hfil $2$\hfil&
          \hfil $1/15$\hfil &\hfil $0$\hfil\\
\hfil$(0,2,0,0)$\hfil&\hfil$-32i$\hfil&
     \hfil $-8\sqrt{3}$\hfil&\hfil $-8$\hfil&
          \hfil $2/3$\hfil &\hfil $3/5$\hfil\\
\hfil$(0,2,0,1)$\hfil&\hfil$-27i$\hfil&
     \hfil $-3\sqrt{3}$\hfil&\hfil $-3$\hfil&
          \hfil $0$\hfil &\hfil $3/80$\hfil\\
\hfil$(0,2,1,0)$\hfil&\hfil$-22i$\hfil&
     \hfil $-8\sqrt{3}$\hfil&\hfil $2$\hfil&
          \hfil $2/3$\hfil &\hfil $3/5$\hfil\\
\hfil$(1,1,0,0)$\hfil&\hfil$-36i$\hfil&
     \hfil $-4\sqrt{3}$\hfil&\hfil $-12$\hfil&
          \hfil $2/5$\hfil &\hfil $1/10$\hfil\\
\hfil$(1,1,0,1)$\hfil&\hfil$-31i$\hfil&
     \hfil $\sqrt{3}$\hfil&\hfil $-7$\hfil&
          \hfil $1/15$\hfil &\hfil $3/80$\hfil\\
\hfil$(1,1,1,0)$\hfil&\hfil$-26i$\hfil&
     \hfil $-4\sqrt{3}$\hfil&\hfil $-2$\hfil&
          \hfil $1/15$\hfil &\hfil $1/10$\hfil\\
\hfil$(2,0,0,0)$\hfil&\hfil$-40i$\hfil&
     \hfil $0$\hfil&\hfil $-16$\hfil&
          \hfil $2/3$\hfil &\hfil $0$\hfil\\
\hfil$(2,0,0,1)$\hfil&\hfil$-35i$\hfil&
     \hfil $5\sqrt{3}$\hfil&\hfil $-11$\hfil&
          \hfil $2/3$\hfil &\hfil $7/16$\hfil\\
\hfil$(2,0,1,0)$\hfil&\hfil$-30i$\hfil&
     \hfil $0$\hfil&\hfil $-6$\hfil&
          \hfil $0$\hfil &\hfil $0$\hfil\\
\hline
\end{tabular}
\renewcommand{\arraystretch}{1.0}
\end{center}
\vspace{.25truecm}

\noindent {\bf Table\ 4.} \ \ \ \ \ Momenta and conformal weights for
 the states $\Vop_{0,(A_0)}$ (\ref{vacA})
 for $p=5,\ q=4$. Note that all momenta in the table have been
 multiplied by a factor $\sqrt{120}$. $h_l$ is the
 contribution of the Liouville sector to the total conformal
 weight $h_{Vir}$.
\vspace{.5truecm}

\vspace{.25cm}
\begin{center}
\renewcommand{\arraystretch}{1.5}
\begin{tabular}{|l||l|l|l|l|l|}
\hline
$(r_1,r_2,t_1,t_2)$ &\hfil $p_2\sqrt{120}$\hfil&\hfil $s_1\sqrt{120}$\hfil
  &\hfil $s_2\sqrt{120}$\hfil
 &\hfil $h_l$\hfil&\hfil $h_{Vir}$\hfil\\
\hline
\hfil$(0,0,0,0)$\hfil&\hfil$-15i$\hfil&
     \hfil $0$\hfil&\hfil $0$\hfil&
          \hfil $0$\hfil &\hfil $7/16$\hfil\\
\hfil$(0,0,1,0)$\hfil&\hfil$-20i$\hfil&
     \hfil $0$\hfil&\hfil $10$\hfil&
          \hfil $2/3$\hfil & \hfil $3/2$\hfil\\
\hfil$(1,0,0,0)$\hfil&\hfil$-11i$\hfil&
     \hfil $0$\hfil&\hfil $-8$\hfil&
          \hfil $1/15$\hfil &\hfil $3/80$\hfil\\
\hfil$(1,0,1,0)$\hfil&\hfil$-16i$\hfil&
     \hfil $0$\hfil&\hfil $2$\hfil&
          \hfil $1/15$\hfil &\hfil $3/5$\hfil\\
\hfil$(2,0,0,0)$\hfil&\hfil$-7i$\hfil&
     \hfil $0$\hfil&\hfil $-16$\hfil&
          \hfil $2/3$\hfil &\hfil $3/80$\hfil\\
\hfil$(2,0,1,0)$\hfil&\hfil$-12i$\hfil&
     \hfil $0$\hfil&\hfil $-6$\hfil&
          \hfil $0$\hfil &\hfil $1/10$\hfil\\
\hline
\end{tabular}
\renewcommand{\arraystretch}{1.0}
\end{center}
\vspace{.25truecm}

\noindent {\bf Table\ 5.} \ \ \ \ \ Momenta and conformal weights for
 the states $\Vop_{1,(A_1)}$ (\ref{stateA})
 for $p=5,\ q=4$. Note that all momenta in the table have been
 multiplied by a factor $\sqrt{120}$. $h_l$ is the
 contribution of the Liouville sector to the total conformal
 weight $h_{Vir}$.
\vspace{.5truecm}

The multiplicity of Virasoro weights $h_{Vir}(r_2,t_2)$ is equal to
\begin{eqnarray}
  (p-2-r_2)(q-2-t_2) + r_2t_2
\end{eqnarray}
The second term is due to the possibility of making the
 transformation (\ref{Virtrans}), and contributes only if $r_2$ and
 $t_2$ are both unequal to zero.

The screening operators which change $r_1$ and $t_1$ (\ref{ScrvacA})
lift part of the degeneracy. The only remaining double multiplicity
occurs for the labels $(0,2,0,1)$ and $(0,1,0,1)$ for weight $3/80$.
This is lifted by the transformation (\ref{ScrvacAA}).

Now consider the states $(A_1)$ at level 1, with $r_2=t_2=0$.
The total number of states is therefore equal to $(p-2)(q-2)$.
For any $(p,q)$ the multiplicity of Virasoro weights is either
1 or 2, depending on the possibility of making the transformation
(\ref{Virtrans}). Multiplicity 2 occurs when
$r_1\ge 1$ and $t_1\le q-4$.
The Virasoro weight which was absent at level 0
occurs for $r_1=0,\ t_1=q-3$, and therefore has multiplicity 1.
In Table 5 we give the states $(A_1)$ at level 1.
Note that indeed the Virasoro weight 0 is absent, and
that the new weight $3/2$ occurs with multiplicity 1. The only
weight with multiplicity 2 is $3/80$. In the previous section we
showed that all states except the one with Virasoro weight $3/2$
can be obtained by screening operators from level 0.

\vspace{0.5truecm}
\noindent{\bf 6.\ Discussion}
\vspace{0.5truecm}


So far we have done all calculations in terms of free scalar fields with
background charges, although we restricted the Liouville momenta to
those corresponding to a $W_3$ minimal model. Let us now discuss what
happens if we compute the cohomology in case the Liouville sector is
a $W_3$ minimal model. To be able to apply the results obtained so far,
we need a Felder resolution describing the irreducible $W_3$ representations
that constitute
the minimal model. This resolution has
been conjectured in \cite{sb}. If we denote the Fock space of $\sigma_1$
and $\sigma_2$, with momenta given by (\ref{s1val}) and (\ref{s2val}),
by ${\cal F}_{r_1,r_2;t_1,t_2}$, then the Felder resolution is
\be \cdots \label{eqq2}
{\cal F}_{-2} \stackrel{Q_{-2}}{\rightarrow}
{\cal F}_{-1} \stackrel{Q_{-1}}{\rightarrow}
{\cal F}_{0} \stackrel{Q_0}{\rightarrow}
{\cal F}_{1} \stackrel{Q_1}{\rightarrow} \cdots
\ee
where $Q_i$ is a sum of products of the screening operators $T_i$, and
\ba
{\cal F}_{2i} & = &  \nonumber
\bigoplus_{k_1+k_2+i=0} {\cal F}_{r_1+(2k_1-k_2)p,r_2+(2k_2-k-1)p;t_1,t_2}
\\ & & \oplus \nonumber
\bigoplus_{k_1+k_2+i=1} {\cal
F}_{r_2+(2k_1-k_2)p,-4-r_1-r_2+(2k_2-k-1)p;t_1,t_2}
\\ & & \oplus
\bigoplus_{k_1+k_2+i=1} {\cal
F}_{-4-r_1-r_2+(2k_1-k_2)p,r_1+(2k_2-k-1)p;t_1,t_2}
\\
{\cal F}_{2i+1} & = &  \nonumber
\bigoplus_{k_1+k_2+i=0} {\cal
F}_{-2-r_1+(2k_1-k_2)p,r_1+r_2+1+(2k_2-k-1)p;t_1,t_2}
\\ & & \oplus \nonumber
\bigoplus_{k_1+k_2+i=0} {\cal
F}_{r_1+r_2+1+(2k_1-k_2)p,-2-r_2+(2k_2-k-1)p;t_1,t_2}
\\ & & \oplus
\bigoplus_{k_1+k_2+i=1} {\cal
F}_{-2-r_1+(2k_1-k_2)p,-2-r_2+(2k_2-k-1)p;t_1,t_2}
\ea
Collectively, we denote this complex by
${\cal F}_{\ast}(r_1,r_2;t_1,t_2)$
The conjecture is that the zeroth cohomology of (\ref{eqq2}) is isomorphic
to an irreducible $W_3$ representation, and all other cohomologies vanish.
We denote this symbolically by ${\cal H}^{\rm min}_{r_1,r_2;t_1,t_2}
=H^{\ast}_Q(
{\cal F}_{\ast}(r_1,r_2;t_1,t_2))$.
An example, which is relevant to the critical $W_3$ string, is to take
the $(4,3)$ $W_3$ minimal model. In this case taking
the cohomology with respect to the BRST operator in (\ref{eqq2}) should be the
same as putting the fields $\sigma_1=\sigma_2=0$ by hand.\\

For a general minimal model, we are interested in
the cohomology of $\Qop_1$ acting on the tensor product of the ghost
Hilbert space ${\cal H}_{\beta,\gamma}$, the Fock space
${\cal H}^{\phi_2}_{p_2}$ and
${\cal H}^{\rm min}_{r_1,r_2;t_1,t_2}$. Using (\ref{eqq2}) this cohomology is
\begin{equation}
H^{\ast}_{\Qop_1} (
{\cal H}_{\beta,\gamma} \otimes
{\cal H}^{\phi_2}_{p_2} \otimes
H^{\ast}_Q({\cal F}_{\ast}(r_1,r_2;t_1,t_2)),
\end{equation}
which, under certain assumptions, is the same as
\begin{equation}
H^{\ast}_{Q} (
H^{\ast}_{\Qop_1} (
{\cal H}_{\beta,\gamma} \otimes
{\cal H}^{\phi_2}_{p_2} \otimes
{\cal F}_{\ast}(r_1,r_2;t_1,t_2))).
\end{equation}
This demonstrates that if the Liouville sector is a $W_3$ minimal model,
we have to drop the states in the $\Qop_1$ cohomology that can be build
out of other states using the screening operators $T_i$ alone. However,
states obtained by acting with the screening operators $R_i$ and $S_i$
represent
new states in the noncritical $W_3$ string. This is an important
observation, as it shows that the noncritical $W_3$ string does not simply
reduce to an ordinary noncritical string theory.
Actually, one starts to suspect that given any realization of a $c=
c_{\rm vir}^{(p,q)}$ Virasoro algebra, it is always possible to build
a generalized Felder complex whose cohomology is precisely that of the
associated Virasoro minimal model. In our case this generalized Felder
complex contains the Fock spaces of $\phi_2,\sigma_1,\sigma_2$ and the
ghost Hilbert space ${\cal H}_{\beta,\gamma}$, and the generalized Felder
BRST operator is composed of $\Qop_1$ and the screening operators
$R_i,S_i,T_i$. It would be interesting to see if one can make this
conjecture more precise.\\

Once we know the $\Qop_1$ cohomology, we can try to use this knowledge to
compute the $\Qop_0+\Qop_1$ cohomology. Assuming the spectral sequence
associated to the decomposition $\Qop=\Qop_0+\Qop_1$ collapses after the
second term, the $\Qop_0+\Qop_1$ cohomology is the same as the $\Qop_0$
cohomology of the $\Qop_1$ cohomology. Now the $\Qop_1$ cohomology is
the direct sum of a set of Virasoro modules with respect to the stress-energy
tensor $T_L+T_{\phi_2}+T_{\beta,\gamma}$.
If we know what type
of Virasoro modules these are, we can use the results for the noncritical
string to find the $\Qop_0$ cohomology on this space,
since $\Qop_0$ is simply the BRST operator of an ordinary
noncritical string theory. Finally, given any state in the $\Qop_0$
cohomology of the $\Qop_1$ cohomology, we can use a standard tic-tac-toe
construction \cite{bot1}
to construct representatives for the $\Qop_0+\Qop_1$ cohomology. It is
straightforward to apply this to the states we computed in previous
sections to obtain some of the known states in the $\Qop_0+\Qop_1$
cohomology.

Additional insight in the structure of the $\Qop_1$ and $\Qop_0$
cohomology can be
obtained by the group-theoretical interpretation of the decomposition
$\Qop=\Qop_0+\Qop_1$ sketched at the end of section~1. There we explained that
the $\Qop_1$ part of $\Qop$ is related to a projection on the space
perpendicular to the first simple root $\alpha_1$.  To see what this implies
for the cohomology, we notice that there is a natural ansatz \cite{bo4}
for the
$\Qop$ cohomology, based on comparison with the ordinary noncritical
string, see also \cite{bo1,bo1a}. This ansatz reads as follows:
represent the ireducible $W_3$ representation labeled by $(r_1,r_2;t_1,t_2)$
by two $sl_3$ weights $\Lambda^{(+)}_{r_1,r_2}=r_1\Lambda_1+r_2
\Lambda_2$ and $\Lambda^{(-)}_{t_1,t_2}=t_1 \Lambda_1+t_2 \Lambda_2$.
As usual, $\alpha_i$ are the simple roots and $\Lambda_i$ the fundamental
weights of $sl_3$. Assume that we are dealing with the $(p,q)$
$W_3$ minimal model, with ${\rm gcd}(p,q)=1$, so that $\alpha=\sqrt{q/p}$.
The matter momenta $p_1,p_2$ can be combined to one weight
\be
\Lambda_M=-i\left(\frac{p_1}{\sqrt{2}}\alpha_1+
p_2\Lambda_2 \sqrt{\frac{3}{2}}\right).
\ee
Let $W$ be the Weyl group, $\hat{W}$ the affine Weyl group and $\rho$ half
the sum of the positive roots, $\rho=\alpha_1+\alpha_2$. Then
the ansatz reads that there is a quartet of states at ghost number
$3+l_w(\hat{w})$, $4+l_w(\hat{w})$, $4+l_w(\hat{w})$ and $5+l_w(\hat{w})$,
if $w\in W$ and $\hat{w}\in\hat{W}$ exist such that
\be \label{eqx}
w^{-1}\left( \alpha \hat{w}\ast\Lambda^{(+)} - \frac{1}{\alpha}
\Lambda^{(-)}+ (\alpha-\frac{1}{\alpha}) \rho \right) =
\left(\Lambda_M+(\alpha+\frac{1}{\alpha})\rho\right).
\ee
Here, $l_w(\hat{w})$ is the twisted length of $\hat{w}\in\hat{W}$ \cite{twil}.
If $\hat{w}=t_{\gamma} w_0$, with $t_{\gamma}$ a translation in the
$\gamma$ direction, $\gamma\in({\bf Z}\alpha_1 + {\bf Z} \alpha_2)$,
and $w_0\in W$, then
\be \hat{w}\ast\Lambda^{(+)}=w_0(\Lambda^{(+)} + \rho) -\rho + p\gamma.
\ee
The level of this quartet of states is given by
\be
l=(\alpha((\hat{w}\ast\Lambda^{(+)}+\Lambda^{(+)} )/2)
-\frac{1}{\alpha} \Lambda^{(-)} +
(\alpha-\frac{1}{\alpha})\rho, \alpha(
\hat{w}\ast\Lambda^{(+)}-\Lambda^{(+)} ) ).
\ee
As a non-trivial check, it has been verified \cite{bo4}
that this correctly reproduces
all states in the critical $W_3$ string \cite{po5},
 by putting $\Lambda^{(+)}= \Lambda^{(-)}=0$ in (\ref{eqx}).
The operators $x$ and $y$ of \cite{po5} are related to particular
translations $t_{\gamma}$ in the root lattice.
If we decompose (\ref{eqx}) in a component in the $\alpha_1$ direction and
in the $\Lambda_2$ direction, we get two equations. The $\Lambda_2$
component determines $p_2$, and should be the equation that describes the
$\Qop_1$ cohomology. The other equation describes the usual noncritical
$\Qop_0$ Virasoro cohomology. A possible proof of (\ref{eqx}) might
consist of separately proving its $\alpha_1$ and $\Lambda_2$ components,
using results for the $\Qop_1$ and $\Qop_0$ cohomology, although this
obscures the group theoretical structure of (\ref{eqx}).
An important message is that by projecting
(\ref{eqx}) onto $\alpha_1$, which reduces the cohomology basically
to that of the noncritical Virasoro string,
we lose information. This information loss is accomplished by the
screening operators $R_i$ and $S_i$, if we use them to identify
states. \\

Finally, let us summarize the main results and ideas put forward in this
paper, generalized to an arbitrary $W_N$ $(p,q)$ minimal model coupled to
$W_N$ gravity. Denote by $\Qop_N$ the BRST operator for this theory.
Then
\begin{itemize}
\item There is a series of BRST operators $\Qop^i_N$, $i=2\ldots N$,
with $\Qop^N_N \subset \cdots \subset \Qop^2_N$, such that $(\Qop^n_N)^2=0$,
and $\Qop_N^n$ involves only $N-n+1$ matter fields, all Liouville fields,
and the ghosts system of spin $n$ up to spin $N$.
\item This decomposition induces a map $H^{\ast}_{\Qop_N}
\rightarrow H^{\ast}_{\Qop^n_N}$ which is neither
surjective nor injective. In the root space of $sl_N$, this
projection is given by a projection on the fundamental weights
$\Lambda_{n-1}\ldots\Lambda_{N-1}$.
\item The cohomology $H^{\ast}_{\Qop^n_N}$ forms a (reducible)
module for the $W_{n-1}$ algebra, and there is a map
$H^{\ast}_{\Qop^n_N} \rightarrow {\cal H}^{p,q}_{min}(W_{n-1})$, where
the latter denotes the Hilbert space of the $(p,q)$ minimal model for
the $W_{n-1}$ algebra. This map is surjective but not injective.
\item There exists an additional set of screening operators acting on
$H^{\ast}_{\Qop^n_N}$. If we identify the states in $H^{\ast}_{\Qop^n_N}$
that are obtained by acting with these new screening operators,
then the previous
map turns into an isomorphism. This yields a new kind of resolution,
different from the usual Felder one, for the $(p,q)$ minimal models of
the $W_{n-1}$ algebra.
\item The noncritical $(p,q)$ $W_{n-1}$ string is a subsector of the
$(p,q)$ noncritical $W_{N}$ string. A correlation function in the
noncritical $W_{N}$ string contains a correlation function
of the noncritical $W_{n}$ string, if one would use the extra screening
operators previously mentioned to cancel the ghost number anomalies
of the ghosts of spin $n\ldots N$. In this way one avoids the problem
of dealing with the $W_N$ moduli. However, as we argued, it is not allowed
to use these screening operators in the $W_{N}$ string, and if the
ghost number anomalies do not match correctly, one still
needs an additional integration over the $W_N$ moduli to compute
these correlation functions. Thus, the $W_{n-1}$ string does not
solve the full $W_N$ string.
\end{itemize}
A more detailed investigation of these statements will be left to
future work.

\vspace{1truecm}

\noindent {\bf Acknowledgements}

\vspace{.5truecm}

The work of E.B.~has been made possible by a fellowship of the Royal
Netherlands Academy of Arts and Sciences (KNAW). Most of the
calculations in this paper have been performed by using the Mathematica
package of \cite{th1} for computing operator product expansions. JdB
is sponsored in part by NSF Grant No. PHY9309888. JdB en TT would
like to thank the Institute for Theoretical Physics, Groningen,
for its hospitality. We would like to thank Peter Bouwknegt for
discussions and several useful comments on an earlier version of the
paper. EB and MdR thank Harm Jan Boonstra for useful discussions.

\vfill\eject

\end{document}